\documentclass[12pt]{article}
\usepackage{amsmath}
\usepackage{amsthm}
\usepackage{amssymb}
\usepackage{mathtools}
\usepackage{graphicx}
\usepackage{enumitem}
\usepackage{natbib}
\usepackage{url} 
\usepackage{booktabs}
\usepackage{color}
\usepackage{array}
\usepackage{multirow}
\usepackage{algorithm}
\usepackage{algpseudocode}
\usepackage{subcaption}
\usepackage{caption}
\usepackage{longtable}

\usepackage{MyArticle}

\allowdisplaybreaks
\def\spacingset#1{\renewcommand{\baselinestretch}%
{#1}\small\normalsize} \spacingset{1}

\newcommand{\blind}{0}

\begin{document}
	
\if0\blind
{
\title{\Large{\textbf{High-dimensional Response Growth Curve Modeling for Longitudinal Neuroimaging Analysis}}}
\author{
Lu Wang \\
\normalsize{\textit{Central South University, Changsha, China}} 
\and		
Xiang Lyu \\
\normalsize{\textit{University of California, Berkeley, USA}} 
\and
Zhengwu Zhang \\
\normalsize{\textit{University of North Carolina, Chapel Hill, USA}} 
\and 
Lexin Li \\
\normalsize{\textit{University of California, Berkeley, USA}}
}
\date{}
\maketitle
} \fi

\if1\blind
{
\title{\Large{\textbf{High-dimensional Response Growth Curve Modeling for Longitudinal Neuroimaging Analysis}}}
\author{
\bigskip
\vspace{1.6in}
}
\date{}
\maketitle
\medskip
} \fi

\begin{abstract}
There is increasing interest in modeling high-dimensional longitudinal outcomes in applications such as developmental neuroimaging research. Growth curve model offers a useful tool to capture both the mean growth pattern across individuals, as well as the dynamic changes of outcomes over time within each individual. However, when the number of outcomes is large, it becomes challenging and often infeasible to tackle the large covariance matrix of the random effects involved in the model. In this article, we propose a high-dimensional response growth curve model, with three novel components: a low-rank factor model structure that substantially reduces the number of parameters in the large covariance matrix, a re-parameterization formulation coupled with a sparsity penalty that selects important fixed and random effect terms, and a computational trick that turns the inversion of a large matrix into the inversion of a stack of small matrices and thus considerably speeds up the computation. We develop an efficient expectation-maximization type estimation algorithm, and demonstrate the competitive performance of the proposed method through both simulations and a longitudinal study of brain structural connectivity in association with human immunodeficiency virus. 
\end{abstract}

\noindent%
{\it Keywords:} Diffusion tensor imaging; Growth curve model; Large covariance matrix; Longitudinal neuroimaging; Low-rank factor model; Sparsity.
\vfill

\newpage
\spacingset{1.5}

\section{Introduction}
\label{sec:introduction}

Longitudinal neuroimaging studies, which repeatedly collect imaging scans of the same subjects over time, are fast emerging in recent years. These studies enable scientists to track the development of brain structures and functions, as well as progression of neurological disorders. The clinical, demographic and genetic data that are often collected concurrently allow scientists to study the influence of environmental and genetic factors on such gradual development and progression  \citep{Madhyastha2018, King2018}. Our motivation example is a longitudinal study of brain structural connectivity in association with human immunodeficiency virus (HIV) \citep{tivarusME2021}. The data consists of diffusion tensor imaging (DTI) of 32 HIV infected patients and 60 age-matched healthy controls over a span of two years. While the healthy subjects were scanned at the baseline then annually, the HIV patients were scanned before starting the treatment of a combination antiretroviral therapy, then after 12 weeks, one year, and two years of the treatment. Each subject in total received DTI scans at 3 to 4 time points, and the brain structural connectivity at each scan was summarized in the form of a 2006-dimensional vector of the fractional anisotropy measures. One of the scientific goals is to investigate the dynamic change of brain structural connectivity and the difference between HIV patients and healthy controls.

Growth curve model (GCM) is a popular approach for studying longitudinal development, and enjoys numerous advantages \citep{curran2012multivariate}. It can characterize the shape and rate of the population-level development, as well as the individual differences in trajectories. It also permits a flexible way of modeling the time variable, so that the data can be collected at uneven time intervals, and each individual can have a varying number of time points. Moreover, GCM for multivariate response can characterize complex correlation structures, through different random effects for different outcomes, including the temporal correlations among the repeated measurements within each outcome, the spatial correlations among different outcomes at a specific time point, and the correlations among outcomes at different time points. On the other hand, the size of the covariance matrix to be estimated increases quadratically with the number of outcome variables, and thus GCM with high-dimensional response faces serious challenges, in terms of both the intensive computation and the limited sample size. 

There have been a number of proposals for modeling multi-response longitudinal data. Notably, \citet{ordaz2013longitudinal} employed the GCM but fitted each response variable one-at-a-time. Although computationally simple, this approach completely ignores the associations among the outcomes. \citet{an2013latent} proposed a latent factor linear mixed model, which reduces the high-dimensional outcomes to some low-dimensional latent factors. The model benefits from dimension reduction, but the resulting latent factors are harder to interpret. Besides, it does not specify how developments of the observed outcomes are correlated. \citet{lu2017bayesian} developed a Bayesian semi-parametric mixed effects model, which characterizes the correlation structure of high-dimensional residuals at a given time point by a sparse factor model \citep{bhattacharya2011sparse}. However, the model assumes that the random effects of different outcomes are independent, and ignores the correlations between subject-specific initial levels and growth rates that are often of key scientific interest. 

There are also some related lines of research. One line concerns sparsity and low-rank factor models for large covariance and precision matrices; see \citet{fan2016overview} for an excellent review. The other line concerns model selection in the context of linear mixed effects model. In particular, \citet{bondell2010joint} re-parameterized the covariance matrix by a modified Cholesky decomposition, so that the shrinkage of a single parameter can eliminate the entire row and column of the covariance. But updating the lower-triangular Cholesky matrix is still computationally intensive when the number of random effects is large. \citet{fan2012variable} employed the group lasso penalty, so that the realizations of a random effect are all in or all out. But they replaced the unknown covariance matrix with a diagonal proxy matrix in model estimation, which completely ignores the correlations among the random effects. \citet{ibrahim2011fixed} and \citet{li2018doubly} applied the group lasso penalty on the Cholesky component matrix of the covariance to encourage elimination of entire rows. But the methods still face the challenge of large matrix inversion in GCM with high-dimensional response.

In this article, we propose a high-dimensional response growth curve model, and develop an efficient expectation-maximization type estimation algorithm. Our proposal involves three key components, a low-rank factor model structure that substantially reduces the number of parameters in the large covariance matrix of random effects, a re-parameterization formulation coupled with an $L_1$ type penalty that selects important fixed and random effect terms, and a computational trick that turns the inversion of a large matrix into the inversion of a stack of small matrices and thus considerably speeds up the computation. Our proposal makes useful contributions in several ways. First, by incorporating the GCM framework, our method is able to explicitly model the longitudinal development of each response at both the population and individual levels, as well as its association with the predictors, while accounting for relatively flexible correlation structures. Such insights are particularly useful in biomedical applications, including child developmental studies, dementia studies, among others. Second, our method is able to select significant fixed and random effect terms, which further facilitates the interpretation, and allows us to focus on the truly important response and predictor variables. Finally, to the best of our knowledge, our method is the first to be capable of simultaneously handling a large number of response variables, ranging from tens to thousands. The techniques we develop are far from simple extensions of the existing solutions.  

We adopt the following notation throughout the article. For a positive integer $d \in \NN_+$, let $[d]$ denote the index set $\{1,2,\ldots,d\}$. For a vector $\ab \in \RR^{d}$, let $\| \ab \|_1, \| \ab \|_2$ denote the vector $L_1$, $L_2$ norm, respectively. For a matrix $\Ab \in \RR^{d_1 \times d_2}$, let $\Ab_{[i,j]}$, $\Ab_{[j,\cdot]}$, $\Ab_{[\cdot,j]}$ denote the $(i,j)$th entry, the $j$th row and the $j$th column of $\Ab$, respectively. Let $|\Ab|, \| \Ab \|_F, | \Ab |_1$ denote the determinant, the Frobenius norm, and the vector $L_1$ norm of $\Ab$, respectively. 

The rest of the article is organized as follows. Section \ref{sec:model} presents our high-dimensional response growth curve model, and Section \ref{sec:estimation} develops the estimation procedure. Section \ref{sec:simulations} reports intensive simulations, and Section \ref{sec:application} revisits the motivating brain structural connectivity example. Section \ref{sec:discussion} concludes the paper with a discussion.

\section{Model}
\label{sec:model}

In this section, we begin with a description of the high-dimensional growth curve model, and the interpretation of the main parameters of interest under this model. We then introduce key model structures to reduce the dimensionality and complexity of the model.

\subsection{Growth curve model}
\label{sec:gcm}

We consider the following growth curve model \citep{hox2014multilevel},
\begin{itemize}
\item[-] Level 1 (within individual):
\begin{equation} 
\begin{aligned} 
y_{ijt} & = \beta_{0ij} + \beta_{1ij}g_{it} + (\bgamma_{j}^{*})^\top \wb_{it} + \varepsilon_{ijt},\\
\bvarepsilon_{it} & = (\varepsilon_{i1t},\dots,\varepsilon_{irt})^{\top}\overset{iid}{\sim}\text{Normal} (\boldsymbol{0},\bSigma^*),\;\; j \in [r],\ t \in [T_{i}];
\end{aligned}
\label{MRGCM-L1}
\end{equation}
	
\item[-] Level 2 (between individual):
\begin{equation}
\begin{aligned}
\beta_{0ij} & = \mu_{0j}^{*} + (\balpha_{0j}^{*})^\top \ub_i + \zeta_{0ij},\\
\beta_{1ij} & = \mu_{1j}^{*} + (\balpha_{1j}^{*})^\top \ub_i + \zeta_{1ij},\\
\bzeta_{i}&  = (\zeta_{0i1},\zeta_{1i1},\ldots,\zeta_{0ir},\zeta_{1ir})^{\top} \overset{iid}{\sim}\text{Normal}\left(\boldsymbol{0},\Gb^* \right),\;\; i \in [n],
\end{aligned}
\label{MRGCM-L2}
\end{equation}	
\end{itemize}
where $y_{ijt}$ denotes the $j$th response variable of the $i$th subject at time point $t$, $g_{it} \in \RR$ is the time variable and $\wb_{it} \in \RR^{p''}$ collects the time-varying predictors of subject $i$ at time $t$, $\ub_i \in \RR^{p'}$ collects the time-invariant predictors of subject $i$, $i \in [n], j \in [r], t \in [T_i]$, and $\varepsilon_{ijt}, \zeta_{0ij}, \zeta_{1ij}$ are the random errors. In our motivation example, $y_{ijt}$ represents the strength of brain structural connectivity, $g_{it}$ is the chronological age, $\ub_i$ is the HIV infection status, and there is no $\wb_{it}$, where $n=92, r=2006$, and $T_{i} \in \{3,4\}$ . 

Models \eqref{MRGCM-L1} and \eqref{MRGCM-L2} involve a set of coefficients related to the mean effect. Among them, $\beta_{0ij}$ and $\beta_{1ij}$ characterize the initial state and the growth rate of the mean growth curve of the $j$th outcome of the $i$th subject, after controlling for the time-varying covariate effect $\bgamma_{j}^{*}$. $\mu_{0j}^{*}$ and $\mu_{1j}^{*}$ characterize the population-level initial state and the growth rate of the mean growth curve of the $j$th outcome, after controlling for the time-invariant covariate effect $\balpha_{0j}^{*}$ and $\balpha_{1j}^{*}$. In our example, in addition to the individual-specific growth curve $(\beta_{0ij}+\beta_{1ij}g)$, we are particularly interested in the mean growth curve $[(\mu_{0j}^*+\alpha_{0j}^*)+(\mu_{1j}^*+\alpha_{1j}^*)g]$ for the population of HIV patients, and the mean growth curve $(\mu_{0j}^* +\mu_{1j}^*g)$ for the healthy controls, for a given outcome at a given age $g_{it} = g$. For instance, if $\alpha_{0j}^*$, $\alpha_{1j}^*$ and $\mu_{1j}^*$ are negative, it implies that, for the $j$th structural connectivity, the HIV patients on average show a lower initial strength and a faster decrease when aging compared to the controls.

Models \eqref{MRGCM-L1} and \eqref{MRGCM-L2} also involve a set of coefficients related to the variation. Among them, $\varepsilon_{ijt}$ captures the measurement error of the outcomes, and $\zeta_{0ij}$, $\zeta_{1ij}$ capture the individual-level variation of growth pattern for the random intercept and slope, respectively. The covariance matrix $\Gb^{*}$ encodes the dependence structure among both $r$ response variables and $T_i$ time points for the same subject. We are generally interested in the dependency between the initial levels of two outcomes $\Cov (\zeta_{0ij},\zeta_{0ij'}) = \Gb^*_{[2j-1,2j'-1]}$, the dependency between the growth rates of two outcomes $\Cov (\zeta_{1ij},\zeta_{1ij'}) = \Gb^*_{[2j,2j']}$, and the dependency between the initial level of one outcome and the growth rate of another outcome $\Cov (\zeta_{0ij},\zeta_{1ij'}) = \Gb^*_{[2j-1,2j']}$, $j, j' \in [r]$. 

Combining models \eqref{MRGCM-L1} and \eqref{MRGCM-L2}, we obtain the following linear mixed effects model,
\begin{equation} \label{MRGCM}
y_{ijt} = \underbrace{\mu_{0j}^{*} + (\balpha_{0j}^{*})^\top \ub_i + (\bgamma_{j}^{*})^\top \wb_{it} +  \mu_{1j}^{*} g_{it} + (\balpha_{1j}^{*})^\top \ub_i g_{it}}_{\bbb_j^\top \xb_{it}} + \underbrace{\zeta_{0ij} + \zeta_{1ij} g_{it}}_{\bzeta_{ij}^\top \zb_{ijt}} + \varepsilon_{ijt},
\end{equation}	
where $\xb_{it}=(1, \ub_i^\top , \wb_{it}^\top, g_{it}, \ub_i^\top g_{it})^\top \in \RR^{p}$, $\zb_{ijt} = (1, g_{it})^\top \in \RR^{2}$, $\bbb_j = (\mu_{0j}^*, (\balpha_{0j}^*)^\top, (\bgamma_{j}^*)^\top, \mu_{1j}^*,$  $(\balpha_{1j}^*)^\top)^\top \in \RR^{p}$ collects all the fixed-effect terms, $\bzeta_{ij} = (\zeta_{0ij}, \zeta_{1ij})^\top \in \RR^{2}$ collects the random-effect terms, and $p = 2 + 2 p' + p''$. Next, stacking the outcome variables together, let $\yb_{it}=(y_{i1t}, \ldots, y_{irt})^\top \in \RR^r$, $\Zb_{it} = \Ib_r \otimes (1,g_{it}) \in \RR^{r \times 2r}$, where $\otimes$ denotes the Kronecker product and $\Ib_r \in \RR^{r \times r}$ is the identity matrix, $\Bb^* \in \RR^{r \times p}$ with the $j$th row equal to $\bbb_j^\top$, and $\bzeta_{i} = (\zeta_{0i1},\zeta_{1i1},\ldots,\zeta_{0ir},\zeta_{1ir})^{\top} \in \RR^{2r}$. We rewrite model \eqref{MRGCM} in the matrix form, 
\begin{equation}\label{eq: regression_form_GCM}
\begin{aligned}
& \yb_{it} = \Bb^*  \xb_{it} + \Zb_{it} \bzeta_{i} + \bvarepsilon_{it}, \\
& \bvarepsilon_{it} \overset{iid}{\sim}\N\left(\boldsymbol{0},\bSigma^{*} \right), \;\; 
 \bzeta_i \overset{iid}{\sim}\N\left(\boldsymbol{0}, \Gb^{*} \right).
\end{aligned}
\end{equation}
where $\bSigma^{*} \in \RR^{r \times r}$, and $\Gb^{*} \in \RR^{2r \times 2r}$.

\subsection{Low-dimensional structures}
\label{sec:dim-reduction}

In this article, we focus on the high-dimensional response GCM scenario where the number of responses $r$ is large, and the number of predictors $p$ is small to moderate. Under this setting, model \eqref{eq: regression_form_GCM} involves a large coefficient matrix $\Bb^* \in \RR^{r \times p}$, two gigantic covariance matrices $\bSigma^{*} \in \RR^{r \times r}$, $\Gb^{*} \in \RR^{2r \times 2r}$, and the corresponding total number of parameters can far exceed the sample size. Next, we introduce some low-dimensional structures, including some factor type model as well as sparsity, to effectively reduce the model dimensionality.

First, we impose that $\bSigma^{*}$ is diagonal, and $\Gb^{*}$ admits a factor model form, in that,
\begin{eqnarray}
\bSigma^{*} & = & \diag(\bsigma^*)=\diag (\sigma_{1}^{*},\dots,\sigma_{r}^{*}), \label{eq: diagonal_epsilon} \\
\Gb^{*} & = & \Qb^*  \Qb^{*\top} + \diag(\bdelta^*), \label{eq: factor2}
\end{eqnarray}
where $\sigma_j^* > 0, j \in [r]$, $\Qb^* \in \RR^{2r \times K}$, $\bdelta^* \in \RR^{2r}$, and $K$ is the reduced rank with $K \ll 2r$. The diagonal structure in \eqref{eq: diagonal_epsilon} substantially simplifies $\bSigma^{*}$, but does not lose too much generality, because the response variables are still correlated through $\Gb^{*}$. The same structure for $\bSigma^{*}$ has also been employed in \citet{laird1987maximum} and \citet{an2013latent}. Meanwhile, the factor model in \eqref{eq: factor2} considerably reduces the number of parameters in $\Gb^{*}$ from $r(2r + 1)$ to $2r(K+1)$. Similar factorization like \eqref{eq: factor2} has been commonly used in modeling large covariance matrices \citep[see, e.g.,][among others]{Fan2008, bartholomew201101}. 

Next, we introduce sparsity to model \eqref{eq: regression_form_GCM}. In the context of high-dimensional longitudinal outcome, there are two types of sparsity, on the mean and the variance, that are of particular interest. For the mean part, it is natural to expect that many outcomes have their corresponding population-level growth curves not varying with time, and it is useful to identify those that do change with time. We thus impose sparsity on $\mu_{1j}^*$ and $\balpha_{1j}^*$, $j \in [r]$, which correspond to the last $(p' + 1)$ columns of the fixed-effect coefficient matrix $\Bb^*$ in \eqref{eq: regression_form_GCM}. Meanwhile, we do not impose sparsity on $\mu_{0j}^*$ and $\balpha_{0j}^*$, $j \in [r]$, as they reflect the initial level of the growth curve, which is allowed to differ across different outcomes. 

For the variance part, again, it is natural to expect that the variances of many outcomes stay constant over time, and it is useful to identify those with time-varying variance patterns. Note that, if $\Gb_{[2j,2j]}^* = 0$, then the variance of the random slope of the $j$th outcome across all subjects becomes zero, i.e.,  $\zeta_{1ij}=0$ for all $i \in [n]$, and correspondingly, $\Var (y_{ijt})$ no longer varies with time $t$. Therefore, we impose sparsity on the entire $(2j)$th row and $(2j)$th column of $\Gb^{*}$, $j \in [r]$. Meanwhile, we do not impose sparsity on the $(2j-1)$th row and $(2j-1)$th column of $\Gb^{*}$, $j \in [r]$, because they control the variance of the random intercept $\zeta_{0ij}$, which is allowed to differ across different outcomes and different subjects. 

To achieve such a sparsity pattern on $\Gb^{*}$, we re-parameterize the factor model \eqref{eq: factor2} as,
\begin{equation} \label{eq: factor}
\Gb^{*}=\diag(\db^*) \Rb^{*} \diag(\db^*),
\end{equation}
where $\Rb^{*}  = \Pb^* \Pb^{*\top} + \Ib_{2r} - \diag (\Pb^* \Pb^{*\top}) \in \RR^{2r \times 2r}$ is the correlation matrix, $\db^* \in \RR^{2r}$ scales the correlation $\Rb^{*}$ to the covariance $\Gb^{*}$, with $(d_{j}^{*})^2 = \delta_{j}^* + (\Qb^* \Qb^{*\top})_{[j,j]}, j \in[2r]$, and $\Pb^* \in \RR^{2r \times K}$ satisfies that $\Qb^{*} =  \diag(\db^*)\Pb^*$, and $\|\Pb_{[j,\cdot]}^*\|_2 < 1$ for $j \in [2r]$ to ensure a valid correlation matrix $\Rb^{*}$. As such, there is a one-to-one mapping between the parameterization $\{\Qb^*, \bdelta^*\}$ in \eqref{eq: factor2} and the parameterization $\{\Pb^*, \db^*\}$ in \eqref{eq: factor}. We then impose sparsity on $\db^*$, so that if $d_{2j}^*=0$, we shrink the entire $(2j)$th row and $(2j)$th column of $\Gb^{*}$, $j \in [r]$. We remark that,  \citet{bondell2010joint} and \citet{chen2003random} employed a similar trick of using a single parameter to control the inclusion or exclusion of an entire row and column of a covariance matrix. However, they used a different re-parameterization, in the form $\Gb^{*}=\diag(\db^*) \bGamma^* \bGamma^{*\top} \diag(\db^*)$, where $\bGamma^*$ is a lower triangular matrix. Such a re-parameterization turns out not suitable for our setting, because when $r$ is large, directly estimating $\bGamma^*$ remains computationally intractable, and may result in a large estimation error. This prompts us to propose the re-parameterization in the form of \eqref{eq: factor} instead. 

Following \eqref{eq: factor}, we also re-parameterize the random effects vector $\bzeta_i$ as $\bzeta_i = \diag(\db^*) \bm{\eta}_i$, with $\bm{\eta}_i \in \RR^{2r}$, $i \in [n]$. Then model \eqref{eq: regression_form_GCM} can be rewritten as, 
\begin{equation}
\label{eq: regression_form}
\begin{aligned}
& \yb_{it} = \Bb^*  \xb_{it} +\Zb_{it} \diag(\db^*) \bm{\eta}_{i} + \bvarepsilon_{it}, \\
& \bvarepsilon_{it} \overset{iid}{\sim}\text{Normal} \left(\boldsymbol{0},\diag(\bsigma^*) \right), \;\;  
\bm{\eta}_i \overset{iid}{\sim} \text{Normal} \left(\boldsymbol{0},\Rb^{*} \right),
\end{aligned}
\end{equation}
where $\Rb^{*}$ is a function of $\Pb^*$ as defined before. We impose sparsity on the last $(p' + 1)$ columns of $\Bb^*$ and the even numbered entries of $\db^*$ in model \eqref{eq: regression_form}.

\section{Estimation}
\label{sec:estimation}

In this section, we propose an expectation-maximization (EM) type algorithm to estimate the parameters in our model. We develop a two-stage approach, where we carry out an unpenalized estimation in the first stage, then feed the estimates into the second stage of penalized estimation. We remark that the unpenalized estimation in the first stage is not simply the penalized estimation in the second stage while setting the penalty parameters to zero. This is because we adopt different parameterizations and also different updating methods in the two estimation stages, and we find this way yields the best empirical performance. We also discuss parameter tuning and computation acceleration.

\subsection{Stage one of unpenalized estimation}

In the first stage, we carry out an unpenalized estimation. We adopt the parameterization under model \eqref{eq: regression_form_GCM} along with the low-dimensional structures \eqref{eq: diagonal_epsilon} and \eqref{eq: factor2}, and the set of parameters to estimate is $\tilde{\btheta} = \{\Qb, \bdelta, \Bb, \bsigma\}$. We adopt the EM alternating optimization approach \citep{bezdek2002some}, where each parameter at the M-step has (conditional) analytic solution, and thus the resulting estimation is efficient. We initialize the algorithm by setting $\Qb^{(0)}, \bdelta^{(0)}, \Bb^{(0)}, \bsigma^{(0)}$, and find the algorithm is not sensitive to the choice of the initial values. We terminate the algorithm when the two consecutive estimates are close enough under a pre-specified tolerance $\epsilon$. We first summarize our estimation procedure in Algorithm  \ref{alg: denseEM}, then discuss the key steps in detail.  

\begin{algorithm}[t!]
\caption{Stage1 of unpenalized estimation.} 
\label{alg: denseEM}
\begin{algorithmic}
\State \textbf{Input:} the data $\big\{ (y_{ijt}, g_{it}, \ub_i, \wb_{it}): i \in [n], j \in [r], t \in [T_i] \big\}$, the initial value $\tilde{\btheta}^{(0)} = \left\{ \Qb^{(0)}, \bdelta^{(0)}, \Bb^{(0)}, \bsigma^{(0)} \right\}$, the reduced dimension $K$, and the tolerance level $\epsilon$.
\State \textbf{Repeat:} 
\State  \quad 1. Conduct the E-step via \eqref{cond-var-mean-zeta} and \eqref{Q-function2}.
\State  \quad 2. Conduct the M-step: 
\State \quad \  (2.1) Obtain $\Qb^{(s+1)}$ and $\bdelta^{(s+1)}$ by alternating between \eqref{opt-Q} and \eqref{opt-Delta} until convergence.
\State \quad \  (2.2) Obtain $\Bb^{(s+1)}$ via \eqref{eq: denseM_B}.
\State \quad \  (2.3) Obtain $\bsigma^{(s+1)}$ via \eqref{eq: denseM_Sigma}.
\State \quad 3. Update the iteration $s=s+1$.
\State \textbf{Until}: the relative changes of $\|\Qb\|_F$, $\|\bdelta\|_2$, $\|\Bb\|_F$, and $\|\bsigma\|_2$ are all smaller than $\epsilon$.		
\State \textbf{Output:} $\hat{\tilde{\btheta}} = \big\{ \tilde{\Qb}, \tilde{\bdelta}, \tilde{\Bb}, \tilde{\bsigma} \big\}$.
\end{algorithmic}
\end{algorithm}

\vspace{0.1in}
\noindent
\textbf{E-step}: 
Given the current parameter estimate, $\tilde{\btheta}^{(s)} = \left\{ \Qb^{(s)}, \bdelta^{(s)}, \Bb^{(s)}, \bsigma^{(s)} \right\}$, the E-step involves finding the conditional distribution of the random effects vector $\bzeta_i \in \RR^{2r}$ given the data and the $Q$-function. The conditional distribution is $\text{Normal} \left(\tilde{\mb}_i(\tilde{\btheta}^{(s)}), \tilde{\bOmega}_i(\tilde{\btheta}^{(s)})\right)$, with 
\begin{equation} \label{cond-var-mean-zeta}
\begin{aligned}
\tilde{\mb}_i(\tilde{\btheta}^{(s)}) & =\tilde{\bOmega}_i(\tilde{\btheta}^{(s)})  \sum_{t=1}^{T_{i}}\Zb_{it}^{\top} \bSigma(\bsigma^{(s)})^{-1} \left(\yb_{it}-\Bb^{(s)}\xb_{it}\right) \in \RR^{2r}, \\
\tilde{\bOmega}_i(\tilde{\btheta}^{(s)}) & =\left( \Gb\left( \Qb^{(s)}, \bdelta^{(s)} \right)^{-1}  + \sum_{t=1}^{T_{i}} \Zb_{it}^{\top} \bSigma(\bsigma^{(s)})^{-1} \Zb_{it} \right)^{-1} \in \RR^{2r \times 2r},
\end{aligned}
\end{equation}
where $\Gb\left( \Qb^{(s)}, \bdelta^{(s)} \right) = \Qb^{(s)} (\Qb^{(s)})^\top + \diag(\bdelta^{(s)})$, and $\bSigma(\bsigma^{(s)}) = \diag(\bsigma^{(s)})$. We present an approach to speed up the computation of $\tilde{\bOmega}_i(\tilde{\btheta}^{(s)})$ in \eqref{cond-var-mean-zeta} in Section \ref{sec:acceleration}. 

Denote the conditional expectation given the data, 
\begin{align*}
\tilde{\bPsi}_i(\tilde{\btheta}^{(s)}) = \EE_{\tilde{\btheta}^{(s)}}\left( \bzeta_i \bzeta_i^\top \mid \{\yb_{it}\} \right) = \tilde{\bOmega}_i(\tilde{\btheta}^{(s)}) + \tilde{\mb}_i(\tilde{\btheta}^{(s)}) \tilde{\mb}_i(\tilde{\btheta}^{(s)})^\top.
\end{align*}
The $Q$-function is of the form,
\begin{align}
Q_n(\tilde{\btheta} & \mid \tilde{\btheta}^{(s)}) = -\dfrac{1}{2}\log|\Gb(\Qb,\bdelta)|-\dfrac{1}{2n}\tr\left( \Gb(\Qb,\bdelta)^{-1}\sum_{i=1}^{n} \tilde{\bPsi}_i(\tilde{\btheta}^{(s)}) \right) -\dfrac{\sum_{i=1}^{n} T_i}{2n}\log|\bSigma(\bsigma)| \nonumber \\
& -\dfrac{1}{2n}\sum_{i=1}^{n}\tr\left( \tilde{\bPsi}_i(\tilde{\btheta}^{(s)})\sum_{t=1}^{T_i} \Zb_{it}^{\top}\bSigma(\bsigma)^{-1}\Zb_{it} \right) +\dfrac{1}{n}\sum_{i=1}^{n}\tilde{\mb}_i(\tilde{\btheta}^{(s)})^\top\sum_{t=1}^{T_i}\Zb_{it}^\top\bSigma(\bsigma)^{-1}(\yb_{it}-\Bb\xb_{it}) \nonumber \\ 
& -\dfrac{1}{2n}\sum_{i=1}^{n}\sum_{t=1}^{T_i}(\yb_{it}-\Bb\xb_{it})^{\top}\bSigma(\bsigma)^{-1}(\yb_{it}-\Bb\xb_{it}). \label{Q-function2}
\end{align}

\vspace{0.1in}
\noindent
\textbf{M-step}: 
The M-step proceeds to update $\tilde{\btheta} = \{\Qb, \bdelta, \Bb, \bsigma\}$ by maximizing the above $Q$-function. We observe that there are analytic forms for the update of $\left\{ \Qb, \bdelta \right\}$ given each other, and for $\Bb$ and $\bsigma$ separately. Let $\left\{ \Qb^{(s)}, \bdelta^{(s)}, \Bb^{(s)}, \bsigma^{(s)} \right\}$ denote the estimate at iteration $s$, and $\left\{ \Qb^{(s, s')}, \bdelta^{(s, s')} \right\}$ the estimate within another iterative procedure at sub-iteration $s'$. 

To update $\Qb^{(s)}$ and $\bdelta^{(s)}$, we take the derivative of the $Q$-function, $Q_n(\tilde{\btheta}\mid\tilde{\btheta}^{(s)})$, with respect to $\Qb$ and $\bdelta$, then employ another iterative procedure. We obtain that, 
\begin{eqnarray} 
\Qb^{(s,s'+1)}  & = & \diag(\bdelta^{(s,s')})^{1/2} \Ub^{(s,s')} \left( \bLambda^{(s,s')} - \Ib_{K} \right)^{1/2}, \label{opt-Q}\\
\hat{\delta}^{(s,s'+1)}_j & = & \left(\dfrac{1}{n}\sum_{i=1}^{n} \tilde{\bPsi}_i(\tilde{\btheta}^{(s)}) - \Qb^{(s,s'+1)}(\Qb^{(s,s'+1)})^{\top}\right)_{[j,j]}, \quad j \in [2r], \label{opt-Delta}
\end{eqnarray}
for iterations $s' = 1, 2, \ldots$, where $\Ub^{(s,s')} \in \RR^{2r \times K}$ and $\bLambda^{(s,s')} \in \RR^{K \times K}$ are the matrix of the leading $K$ eigenvectors and the diagonal matrix of the leading $K$ eigenvalues, respectively, of the intermediate matrix, 
\begin{align*}
\Sbb = \diag(\bdelta^{(s,s')})^{-1/2} \left( \dfrac{1}{n}\sum_{i=1}^{n} \tilde{\bPsi}_i(\tilde{\btheta}^{(s)}) \right) \diag(\bdelta^{(s,s')})^{-1/2}.
\end{align*}
To avoid sign flip, we fix the signs of the entries in the first row of $\Ub^{(s,s')} $ in \eqref{opt-Q} to be positive. We begin with $\{ \Qb^{(s)}, \bdelta^{(s)} \}$, iterate through $s'$ until the two consecutive estimates are close enough, and obtain the updated estimate $\{ \Qb^{(s+1)}, \bdelta^{(s+1)} \}$.

To update $\Bb^{(s)}$ and $\bsigma^{(s)}$, we take the derivative of the $Q$-function with respect to $\Bb$ and $\bsigma$, and obtain that, 
\begin{eqnarray}
\Bb^{(s+1)} & = & \left( \sum_{i=1}^n \sum_{t=1}^{T_i}  \left(\yb_{it} - \Zb_{it}\tilde{\mb}_{i}(\tilde{\btheta}^{(s)}) \right)   \xb_{it}^\top \right) \left( \sum_{i=1}^n \sum_{t=1}^{T_i} \xb_{it}  \xb_{it}^\top \right)^{-1} \label{eq: denseM_B} \\
\sigma^{(s+1)}_j & = & \frac{1}{\sum_{i=1}^{n} T_i} \sum_{i=1}^n \sum_{t=1}^{T_i} \bigg( \left( \Zb_{it} \tilde{\bPsi}_{i}(\tilde{\btheta}^{(s)})  \Zb_{it}^\top \right)_{[j,j]}  + \left( \yb_{it}- \Bb^{(s+1)}  \xb_{it} \right)_{[j]}^2 \nonumber \\ 
& & \hskip3cm - 2 \left( \yb_{it}- \Bb^{(s+1)}  \xb_{it} \right)_{[j]} \left( \Zb_{it}\tilde{\mb}_{i}(\tilde{\btheta}^{(s)}) \right)_{[j]} \bigg), \, j \in [r]. 
\label{eq: denseM_Sigma}		
\end{eqnarray}

\subsection{Stage two of penalized estimation}

In the second stage, we carry out a penalized estimation. To facilitate the implementation of the penalty, we adopt the parameterization under model \eqref{eq: regression_form} along with the low-dimensional structures \eqref{eq: diagonal_epsilon} and \eqref{eq: factor}, and the set of parameters to estimate becomes $\btheta = \{\Pb, \db, \Bb, \bsigma\}$. We adopt the projected gradient descent \citep{Beck2017} and coordinate descent \citep{friedman2010regularization} for parameter estimation. We initialize this stage by adopting the estimate $\hat{\tilde{\btheta}} = \big\{ \tilde{\Qb}, \tilde{\bdelta}, \tilde{\Bb}, \tilde{\bsigma} \big\}$ from the first stage as the starting values, i.e., 
\vspace{-0.01in}
\begin{equation} \label{eqn:initial}
\Pb^{(0)} = \diag(\db^{(0)})^{-1} \tilde{\Qb}; \;\; 
d^{(0)}_j = \sqrt{ (\tilde{\Qb} \tilde{\Qb}^\top)_{[j,j]} + \tilde{\delta}_j }, \ j \in [2r]; \;\; 
\Bb^{(0)} = \tilde{\Bb}; \;\; 
\bsigma^{(0)} = \tilde{\bsigma}.
\end{equation}
We again terminate the algorithm when the two consecutive estimates are close. We first summarize our estimation procedure in Algorithm  \ref{alg: sparseEM}, then discuss the key steps in detail.  

\begin{algorithm}[t!]
\caption{Stage 2 of penalized estimation.} 
\label{alg: sparseEM}
\begin{algorithmic}
\State \textbf{Input:} the data $\big\{ (y_{ijt}, g_{it}, \ub_i, \wb_{it}): j\in [r], t\in [T_i], i\in [n] \big\}$, the initial value $\btheta^{(0)} = \{ \Pb^{(0)}, \db^{(0)}, \Bb^{(0)}, \bsigma^{(0)} \}$ from \eqref{eqn:initial}, the reduced dimension $K$, and the tolerance level $\epsilon$.
\State \textbf{Repeat:} 
\State  \quad 1. Conduct the E-step via \eqref{cond-var-mean-eta} and \eqref{Q-function}.		
\State  \quad 2. Conduct the sparse M-step:
\State \quad \ (2.1) Obtain $\Pb^{(s+1)}$ via \eqref{eq: M-P}. 
\State \quad \ (2.2) \textbf{Repeat:}
\State \quad \quad \ (2.2.1) Obtain $d_{2j-1}^{(s,s'+1)}$ via \eqref{eq: sparseM_d1}, and $d_{2j}^{(s,s'+1)}$ via \eqref{eq: sparseM_d2_solu}, $j\in[r]$.
\State \quad \quad \ (2.2.2) Obtain the first $(p - p' - 1)$ columns of $\Bb^{(s,s'+1)}$ via \eqref{eq: sparseM_B1}, and the last $(p'+1)$ \\
\quad \quad \quad \quad \quad \, columns of $\Bb^{(s,s'+1)}$ via \eqref{eq: sparseM_B2}.		
\State \quad \quad \ (2.2.3) Compute $\bsigma^{(s,s'+1)}$ via \eqref{eq: M_epsilon}. 
\State \quad \quad \  \textbf{Until} convergence.		
\State \quad 3. Update the iteration $s=s+1$.  
\State \textbf{Until}: the relative changes of $\|\Pb\|_F$, $\|\db\|_2$, $\|\Bb\|_F$, and $\|\bsigma\|_2$ are all smaller than $\epsilon$.
\State \textbf{Output:} $\hat{\btheta} = \big\{ \hat{\Pb}, \hat{\db}, \hat{\Bb}, \hat{\bsigma} \big\}$. 
\end{algorithmic}
\end{algorithm}

\vspace{0.1in}
\noindent
\textbf{E-step}: 
Given the current parameter estimate, $\btheta^{(s)} = \left\{ \Pb^{(s)}, \db^{(s)}, \Bb^{(s)}, \bsigma^{(s)} \right\}$, the conditional distribution of $\bm{\eta}_i \in \RR^{2r}$ given the data is $\text{Normal} \left(\mb_i(\btheta^{(s)}), \bOmega_i(\btheta^{(s)})\right)$, with
\begin{eqnarray}
\mb_i(\btheta^{(s)}) & = & \bOmega_i(\btheta^{(s)}) \diag(\db^{(s)}) \sum_{t=1}^{T_{i}}\Zb_{it}^{\top}\bSigma(\bsigma^{(s)})^{-1} \left(\yb_{it}-\Bb^{(s)}\xb_{it}\right) \in \RR^{2r}, \label{cond-var-mean-eta} \\
\bOmega_i(\btheta^{(s)}) & = & \left( \Rb(\Pb^{(s)})^{-1} + \diag(\db^{(s)}) \Big( \sum_{t=1}^{T_{i}} \Zb_{it}^{\top}\bSigma(\bsigma^{(s)})^{-1}\Zb_{it} \Big) \diag(\db^{(s)}) \right)^{-1} \in \RR^{2r \times 2r}, \nonumber
\end{eqnarray}
where $\Rb(\Pb^{(s)}) = \Pb^{(s)} (\Pb^{(s)})^\top + \Ib_{2r} - \diag (\Pb^{(s)} (\Pb^{(s)})^\top)$, and $\bSigma(\bsigma^{(s)}) = \diag(\bsigma^{(s)})$. We again discuss how to speed up the computation of $\bOmega_i(\btheta^{(s)})$ in \eqref{cond-var-mean-eta} in Section \ref{sec:acceleration}. 

Denote the conditional expectation given the data, 
\begin{align*}
\bPsi_i(\btheta^{(s)}) = \EE_{\btheta^{(s)}}\left( \bm{\eta}_i \bm{\eta}_i^\top \mid \{\yb_{it}\} \right) = \bOmega_i(\btheta^{(s)}) + \mb_i(\btheta^{(s)}) \mb_i(\btheta^{(s)})^\top.
\end{align*} 
The $Q$-function is of the form, 
\begin{align}
Q_n(\btheta \mid \btheta^{(s)}) = & -\dfrac{1}{2}\log|\Rb(\Pb)|-\dfrac{1}{2n}\tr\left( \Rb(\Pb)^{-1}\sum_{i=1}^{n} \bPsi_i(\btheta^{(s)}) \right) -\dfrac{\sum_{i=1}^{n} T_i}{2n}\log|\bSigma(\bsigma)| \nonumber \\
& - \dfrac{1}{2n}\sum_{i=1}^{n}\tr\left( \diag(\db) \Big(\sum_{t=1}^{T_i} \Zb_{it}^{\top}\bSigma(\bsigma)^{-1}\Zb_{it}\Big) \diag(\db) \bPsi_i(\btheta^{(s)}) \right) \nonumber \\
& + \dfrac{1}{n}\sum_{i=1}^{n}\mb_i(\btheta^{(s)})^\top \diag(\db) \sum_{t=1}^{T_i}\Zb_{it}^\top\bSigma(\bsigma)^{-1}(\yb_{it}-\Bb\xb_{it}) \nonumber \\
& - \dfrac{1}{2n}\sum_{i=1}^{n}\sum_{t=1}^{T_i}(\yb_{it}-\Bb\xb_{it})^{\top}\bSigma(\bsigma)^{-1}(\yb_{it}-\Bb\xb_{it}). 	\label{Q-function} 
\end{align}

\vspace{0.1in}
\noindent
\textbf{Sparse M-Step}: 
The M-step proceeds to update $\btheta = \{\Pb, \db, \Bb, \bsigma\}$ by maximizing the above $Q$-function, while in this stage under an additional adaptive $L_1$ penalty. We observe that the estimation of $\Pb$ can be carried out separately from the rest of parameters, and we alternatively update $\{\db, \Bb, \bsigma\}$ in another iterative procedure. Let $\{\Pb^{(s)}, \db^{(s)}, \Bb^{(s)}, \bsigma^{(s)}\}$ denote the estimate at iteration $s$, and $\{\db^{(s, s')}, \Bb^{(s, s')}, \bsigma^{(s, s')}\}$ the estimate within another iterative procedure at sub-iteration $s'$. 

To update $\Pb^{(s)}$, we solve the following optimization using the projected gradient descent algorithm \citep{Beck2017}, 
\begin{equation} \label{eq: M-P}
\begin{aligned}
\Pb^{(s+1)} = \argmin_{\Pb \in \RR^{2r \times K }} \log |\Rb(\Pb)| + \tr\left( \Rb(\Pb)^{-1}\dfrac{1}{n}\sum_{i=1}^n \bPsi_i(\btheta^{(s)}) \right), \textrm{ subject to } \|\Pb_{[j,\cdot]}\|_2 <1, j \in [2r]. 
\end{aligned}
\end{equation}

To update $\db^{(s)}, \Bb^{(s)}$, and $\bsigma^{(s)}$, we employ another iterative procedure. 

To update $\db^{(s)}$, recall that, in model \eqref{eq: regression_form}, we only impose sparsity on the even numbered entries of $\db^*$, so to penalize the variances of the random slopes, but not the random intercepts. Therefore, maximizing the $Q$-function, $Q_n(\btheta\mid\btheta^{(s)})$, with respect to $d_{2j-1}$, $j\in[r]$, we obtain that, 
\begin{align} \label{eq: sparseM_d1}
d^{(s,s'+1)}_{2j-1} = \dfrac{ \sum_{i=1}^n  \sum_{t=1}^{T_i} \left( \left(\yb_{it}- \Bb^{(s,s')}  \xb_{it} \right)_{[j]} \mb_{i}(\btheta^{(s)})_{[2j-1]} - d^{(s,s')}_{2j}g_{it}    \bPsi_{i}(\btheta^{(s)})_{[2j-1,2j]} \right) }{ \sum_{i=1}^n T_i \bPsi_{i}(\btheta^{(s)})_{[2j-1,2j-1]} }.
\end{align}
Meanwhile, maximizing the $Q$-function with respect to $d_{2j}$, $j\in[r]$, under an additional adaptive $L_1$ penalty, amounts to solving the minimization problem,  
\begin{align*} 
d^{(s,s'+1)}_{2j} = \argmin_{d_{2j} \in \RR} \dfrac{1}{n\sigma^{(s,s')}_j} \sum_{i=1}^n  \sum_{t=1}^{T_i} \bigg( & d_{2j}^2 g_{it}^2  \bPsi_{i}(\btheta^{(s)})_{[2j,2j]} - 2 d_{2j}  g_{it} \Big( (\yb_{it}- \Bb^{(s,s')}  \xb_{it} )_{[j]} \mb_{i}(\btheta^{(s)})_{[2j]} \\
& - d_{2j-1}^{(s,s'+1)}  \bPsi_{i}(\btheta^{(s)})_{[2j-1,2j]} \Big)  \bigg) + \lambda_d \dfrac{| d_{2j} |} {| \bar{d}_{2j}^{(s,s'+1)} |},
\end{align*}
where $\bar{d}_{2j}^{(s,s'+1)}$ is the unpenalized ordinary least squares (OLS) estimate, and $\lambda_d$ is a penalty parameter. Since this minimization is done for one $j$ at a time, we obtain the closed-form solution that, 
\begin{equation} \label{eq: sparseM_d2_solu}
d^{(s,s'+1)}_{2j}= \dfrac{1}{a_j}\mbox{sign}(c_j)\left(|c_j| - \dfrac{\lambda_d} {| \bar{d}_{2j}^{(s,s'+1)} |} \right)_+
\end{equation}
where
\begin{align*}
a_j = & \dfrac{2}{n\sigma^{(s,s')}_j} \sum_{i=1}^n  \sum_{t=1}^{T_i} g_{it}^2  \bPsi_{i}(\btheta^{(s)})_{[2j,2j]} \\
c_j = & \dfrac{2}{n\sigma^{(s,s')}_j} \sum_{i=1}^n  \sum_{t=1}^{T_i}  g_{it} \left( (\yb_{it}- \Bb^{(s,s')}  \xb_{it} )_{[j]} \mb_{i}(\btheta^{(s)})_{[2j]} - d_{2j-1}^{(s,s'+1)}  \bPsi_{i}(\btheta^{(s)})_{[2j-1,2j]} \right) . 
\end{align*}

To update $\Bb^{(s)}$, recall that, in model \eqref{eq: regression_form}, we impose sparsity on the last $(p' + 1)$ columns of $\Bb^*$, so to penalize the fixed effects related to the time variable $g_{it}$ only. Therefore, we split $\Bb^{(s,s')}$ into two parts as $\Bb^{(s,s')} = \left( \Bb_1^{(s,s')}, \Bb_2^{(s,s')} \right)$, where $\Bb_1^{(s,s')} \in \RR^{r \times (p - p' - 1)}$ consists of the first $(p - p' - 1)$ columns of $\Bb$, and $\Bb_2^{(s,s')} \in \RR^{r \times (p' + 1)}$ the last $(p'+1)$ columns of $\Bb$. Correspondingly, we partition the predictor vector $\xb_{it} \in \RR^p$ as $\xb_{it}= (\xb_{it,1}^\top,\xb_{it,2}^\top)^\top$, where $\xb_{it,1}=(1, \ub_i^\top, \wb_{it}^\top)^\top \in \RR^{p - p' - 1}$, and $\xb_{it,2}=(g_{it}, \ub_i^\top g_{it})^\top \in \RR^{p'+1}$. 

For $\Bb_1^{(s,s')}$, we obtain that, 
\begin{equation} \label{eq: sparseM_B1}
\Bb_1^{(s,s'+1)}= \left( \sum_{i=1}^n \sum_{t=1}^{T_i}   \left(\yb_{it}- \Bb_2^{(s,s')}  \xb_{it,2} - \Zb_{it} \diag(\db^{(s,s'+1)})\mb_{i}(\btheta^{(s)}) \right)   \xb_{it,1}^\top \right) \left( \sum_{i=1}^n \sum_{t=1}^{T_i} \xb_{it,1}  \xb_{it,1}^\top \right)^{-1}.
\end{equation}

For $\Bb_2^{(s,s')}$, maximizing the $Q$-function with respect to $\Bb_2$, under an additional adaptive $L_1$ penalty, amounts to solving the minimization problem,  
\begin{align}  
\Bb_2^{(s,s'+1)} = & \min_{\Bb_2 \in \RR^{r \times (p'+1)}} \dfrac{1}{2n}  \sum_{i=1}^n \sum_{t=1}^{T_i} \bigg( ( \Bb_2  \xb_{it,2} )^\top \bSigma(\bsigma^{(s,s')})^{-1} \Bb_2  \xb_{it,2} \nonumber \\
& - 2 \left(\yb_{it}- \Bb_1^{(s,s'+1)}  \xb_{it,1}   - \Zb_{it} \diag(\db^{(s,s'+1)}) \mb_{i}(\btheta^{(s)})\right)^\top \bSigma(\bsigma^{(s,s')})^{-1}  \Bb_2  \xb_{it,2} \bigg) \nonumber \\  
& + \lambda_B  \left| \Bb_2 \oslash \bar{\Bb}_2^{(s,s'+1)} \right|_1, \label{eq: sparseM_B2}
\end{align}
where $\oslash$ is the element-wise division, $\bar{\Bb}_2^{(s,s'+1)}$ is the unpenalized OLS estimate, and $\lambda_B$ is a penalty parameter. We solve \eqref{eq: sparseM_B2} using the coordinate descent \citep{friedman2010regularization}.
	
To update $\bsigma^{(s)}$, we obtain that for $j \in [r]$, 
\begin{align} 
\sigma^{(s,s'+1)}_j & =  \dfrac{1}{\sum_{i=1}^{n} T_i} \sum_{i=1}^n \sum_{t=1}^{T_i} \bigg( \left( \Zb_{it} \diag(\db^{(s,s'+1)}) \bPsi_{i}(\btheta^{(s)}) \diag(\db^{(s,s'+1)}) \Zb_{it}^\top \right)_{[j,j]}    \nonumber \\
& +\left( \yb_{it}- \Bb^{(s,s'+1)}  \xb_{it} \right)_{[j]}^2 - 2\, \left( \yb_{it}- \Bb^{(s,s'+1)}  \xb_{it} \right)_{[j]}\, \left( \Zb_{it}\diag(\db^{(s,s'+1)})\mb_{i}(\btheta^{(s)})\right)_{[j]} \bigg). 
\label{eq: M_epsilon} 	 	
\end{align}

\subsection{Parameter tuning}
\label{sec:tuning}

There are three tuning parameters in our estimation procedure, the reduced rank $K$, and two sparsity penalty parameters $\lambda_d, \lambda_B$. We propose to tune these parameters using Bayesian information criterion (BIC). That is, we minimize the following criterion,
\begin{equation*}
\textrm{BIC}  = -2 \, \ell(\hat{\btheta} | \yb) + \log(n) \, df,
\end{equation*}
where $\ell(\hat{\btheta} | \yb)$ is the log-likelihood function evaluated at the estimate $\hat{\btheta}$, and $df$ is the degree of freedom. We discuss how to speed up the computation of $\ell(\hat{\btheta} | \yb)$ in Section \ref{sec:acceleration}. 

For $K$, we use the output of the unpenalized estimate from the first stage for $\hat{\btheta}$. This way, it avoids tuning all three parameters together, which can be computationally expensive. The degree of freedom $df = 2r(K+1) - K(K-1)/2$, as \eqref{opt-Q} introduces $K(K-1)/2$ constraints to ensure a unique solution for $\Qb$. 

For $\lambda_d, \lambda_B$, we adopt the strategy of \citet{cai2019chime}, and tune them at each iteration, which helps improve both the estimation and selection accuracy empirically. We thus use the estimate at each iteration $\btheta^{(s)} = \left\{ \Pb^{(s)}, \db^{(s)}, \Bb^{(s)}, \bsigma^{(s)} \right\}$ for $\hat{\btheta}$. In particular, for $\lambda_d$, the degree of freedom $df$ is the number of nonzero entries in $\left\{ d^{(s)}_{2j}: j \in [r] \right\}$ times $(K+1)$. Recall the parameter mapping from $(\Pb, \db)$ to $(\Qb, \bdelta)$: $\Qb =  \diag(\db)\Pb, \quad \delta_{j}  = d_j^2 - (\Qb \Qb^{\top})_{[j,j]}, \ j \in[2r]$. Note that a zero $d_j$ results in a zero row in $\Qb$ and a zero $\delta_j$, and thus a decrease of $(K+1)$ in the degree of freedom. For $\lambda_B$, the degree of freedom $df$ is the number of nonzero entries in $\Bb_2^{(s)}$.

\subsection{Computation acceleration}
\label{sec:acceleration}

In the proposed estimation procedure, the E-step involves the inversion of some $2r \times 2r$ matrices in \eqref{cond-var-mean-zeta} and \eqref{cond-var-mean-eta} for each subject $i$, $i \in [n]$. In addition, the log-likelihood function involves the determinant and inversion of some $rT_i \times rT_i$ matrix for each subject $i$, $i \in [n]$. When $r$ is large, e.g., in thousands, these steps can be computationally expensive. We next discuss how to speed up the computations. 

We first discuss the computation of $\bOmega_i(\btheta)$ in \eqref{cond-var-mean-eta}, while the computation of $\tilde{\bOmega}_i(\tilde{\btheta})$ in \eqref{cond-var-mean-zeta} is done similarly. Specifically, since $\bSigma(\bsigma)$ is a diagonal matrix, we have that,
\begin{equation*}
\sum_{t=1}^{T_i} \Zb_{it}^{\top}\bSigma(\bsigma)^{-1}\Zb_{it} = \bSigma(\bsigma)^{-1} \otimes \Ab_i,\ \ \mbox{where} \ \Ab_i = \sum_{t=1}^{T_i} (1,g_{it})^\top (1,g_{it}) \in \RR^{2 \times 2}.
\end{equation*}
Repeatedly applying the Woodbury formula \citep{higham2002wood} and plugging in $\big(\bSigma(\bsigma)^{-1} \otimes \Ab_i \big)^{-1} = \bSigma(\bsigma) \otimes \Ab_i^{-1}$ yields that, 
\begin{equation*}
\bOmega_{i}(\btheta) = \Rb(\Pb) - \Rb(\Pb) \diag(\db) \Big( \Cb_{i}-\Cb_{i} \diag(\db) \Pb\Fb_{i}\Pb^{\top} \diag(\db) \Cb_{i} \Big) \diag(\db) \Rb(\Pb),
\end{equation*}
where $\Cb_{i} = \Big( \bSigma(\bsigma) \otimes \Ab_{i}^{-1} + \diag(\db) \big(\Ib_{2r} - \diag (\Pb \Pb^\top)\big) \diag(\db) \Big)^{-1} \in \RR^{2r \times 2r}$, and $\Fb_i = \Big(\Ib_{K}+ \Pb^{\top} \diag(\db) \Cb_{i} \diag(\db) \Pb \Big)^{-1} \in \RR^{K \times K}$. Since $\bSigma(\bsigma)$, $\diag(\db)$, and $\big(\Ib_{2r} - \diag (\Pb \Pb^\top)\big)$ are all diagonal matrices, the computation of $\Cb_i$ amounts to inverting $r$ $2 \times 2$ matrices, which is computationally simple. 

We next discuss the computation of the log-likelihood function, which is of the form, 
\begin{equation*}
\ell (\btheta | \yb) = -\dfrac{1}{2}\sum_{i=1}^n \log |\Vb_i| - \dfrac{1}{2}\sum_{i=1}^n \Big( \yb_i - (\Ib_{T_i} \otimes \Bb)\xb_i \Big)^\top \Vb_i^{-1} \Big( \yb_i - (\Ib_{T_i} \otimes \Bb)\xb_i \Big), 
\end{equation*}
up to some constant, where $\yb_i \in \RR^{r T_i}$, $\xb_i \in \RR^{p T_i}$, and $\Zb_i \in \RR^{r T_i \times 2r}$ are obtained by stacking $\yb_{it}$, $\xb_{it}$ and $\Zb_{it}$ across all $t \in [T_i]$, respectively, and 
\begin{align*}
\Vb_i = \Zb_i \diag(\db) \Rb(\Pb) \diag(\db) \Zb_i^\top + \Ib_{T_i} \otimes \bSigma(\bsigma) \ \in \RR^{rT_i \times rT_i}.
\end{align*}
Repeatedly applying the Sylvester's determinant theorem \citep{harville2008matrix} yields that, 
\begin{align*}
|\Vb_i| = |\bSigma(\bsigma)|^{T_i-2}|\Ab_i|^r |\Cb_i|^{-1} |\Fb_i|^{-1},
\end{align*}
where $\Ab_i$, $\Cb_i$ and $\Fb_i$ are as defined earlier, whose determinants are easy to compute. Moreover, applying the the Woodbury formula yields that, 
\begin{align*}
\Vb_i^{-1} = \Ib_{T_i} \otimes \bSigma(\bsigma)^{-1} - \big(\Ib_{T_i} \otimes \bSigma(\bsigma)^{-1} \big)\Zb_i \diag(\db) \bOmega_{i}(\btheta) \diag(\db) \Zb_i^\top \big(\Ib_{T_i} \otimes \bSigma(\bsigma)^{-1}\big).
\end{align*}
Then the second term in the log-likelihood function $\ell (\btheta | \yb)$ becomes
\begin{equation*}
- \dfrac{1}{2}\sum_{i=1}^n \sum_{t=1}^{T_i}(\yb_{it}-\Bb\xb_{it})^{\top} \bSigma(\bsigma)^{-1} (\yb_{it}-\Bb\xb_{it}) + \dfrac{1}{2}\sum_{i=1}^n \sum_{t=1}^{T_i} (\yb_{it}-\Bb\xb_{it})^\top \bSigma(\bsigma)^{-1} \Zb_{it} \diag(\db) \mb_i(\btheta),
\end{equation*}
which is straightforward to compute.

\section{Simulations}
\label{sec:simulations}

In this section, we evaluate the empirical performance of the proposed method. We also compare with the alternative solution of fitting one response at a time.

\subsection{Data generation}

We simulate the data following the model setup in \eqref{MRGCM-L1} and \eqref{MRGCM-L2}. We simulate $r$ outcome variables, with four different types. Among them, 70\% outcomes have a constant mean growth curve and a constant variance over time, where $\mu_{1j}^*=0$, and the $2j$th row and $2j$th column of $\Gb^*$ are zero for those $j$s; 10\% outcomes have a time-varying mean growth curve and a constant variance over time, where $\mu_{1j}^*\sim \textrm{Uniform}(1,2)$, and the $2j$th row and $2j$th column of $\Gb^*$ are zero; 10\% outcomes have a constant mean growth curve and a time-varying variance over time, where $\mu_{1j}^*=0$; and 10\% outcomes have a time-varying mean growth curve and a time-varying variance over time, where $\mu_{1j}^* \sim \textrm{Uniform}(-2,-1)$. We generate the nonzero entries in $\Gb^*$ from the factorization $\Qb^*  \Qb^{*\top} + \diag(\bdelta^*)$, where the rank is set at $K^* = 3$, the entries of $\Qb^*$ are sampled from $\textrm{Uniform}(-1,1)$, and $\bdelta^*=(1, \ldots, 1)^\top$. We generate $\gamma_{j}^* \sim \textrm{Normal}(0,0.1^2)$, $\mu_{0j}^* \sim \textrm{Normal}(0,1)$, $\alpha_{0j}^*\sim \textrm{Normal}(0,0.1^2)$, and $\alpha_{1j}^*\sim \textrm{Uniform}(1,2)$ for $5\%$ outcomes, and set $\alpha_{1j}^*=0$ for the rest. For each subject $i$, we set $g_{i1} \sim \textrm{Uniform}(20,60)$, and sample the number of time points $T_i$ randomly from $\{3,4,5\}$. We generate the time-invariant covariate $u_{i} \sim \textrm{Bernoulli}(0.5)$, and the time-varying covariate $w_{it}$ from an AR-1 process.  

We consider four combinations of the response dimension $r$, the sample size $n$, and the noise level that is set as a percentage of the standard deviation of the conditional mean of outcome $j$, including $r=100, n=100, 20\%$ noise, $r=200, n=100, 20\%$ noise, $r=100, n=50, 20\%$ noise, and $r=100, n=100, 50\%$ noise.

\subsection{Estimation and selection accuracy}

We apply the proposed method to the simulated data. We standardize both $g_{it}$ and each covariate of $\xb_{it}$ to have zero mean and unit variance. We initialize the algorithm with $\Qb = \mathbf{0}, \bdelta = \mathbf{1}$. $\mu_{0j}$ and $\sigma_j$ are initialized at the sample mean and sample variance of the $j$th outcome for $j \in [r]$. The other entries of $\Bb$ are initialized at zero. We set the tolerance level at $\epsilon = 0.001$. We also compare with the alternative baseline method of fitting a univariate response GCM one at a time, using restricted maximum likelihood estimation. We report both the parameter estimation accuracy, and the variable selection accuracy in terms of the true positive rate (TPR) and false positive rate (FPR). We also report the percentage of times when the rank $K^*$ is correctly selected. 

\begin{table}[t!]
\centering
\caption{Simulation results over 100 data replications. The evaluation criteria include $\| \hat{\Bb} - \Bb^* \|_{F}^{2}/pr$ - the estimation error of $\Bb$, $\| \hat{\Gb} - \Gb^* \|_{F}^{2}/(2r)^{2}$ - the  estimation error of $\Gb$, ``TPR - fixed" - the proportion of true nonzero fixed slopes in $\Bb^*$ correctly identified, ``FPR - fixed" - the proportion of true zero fixed slopes incorrectly identified, ``TPR - random" - the proportion of true nonzero variances of random slopes correctly identified, ``FPR - random" - the proportion of true zero variances of random slopes incorrectly identified, and ``\% of $\hat{K}=K^*$'' - the percentage of $K^*$ correctly selected.}
\label{table-simu-res}
\footnotesize{ 
\begin{tabular}{>{\centering}m{3cm}|>{\centering}m{2.8cm}|>{\centering}m{2.8cm}|>{\centering}m{2.8cm}|>{\centering}m{2.8cm}}
                \toprule 
                \multirow{2}{2.8cm}{} & $r=100$, $n=100$, 20\% noise & $r=200$, $n=100$, 20\% noise & $r=100$, $n=50$, 20\% noise & $r=100$, $n=100$, 50\% noise \tabularnewline
                \cline{2-5} 
                 & \multicolumn{4}{c}{Proposed high-dimensional response GCM method }\tabularnewline
                \midrule 
                $\| \hat{\Bb} - \Bb^* \|_{F}^{2}/pr$ & 0.0179 $\pm$ 0.0029 & 0.0186 $\pm$ 0.0030 & 0.0408 $\pm$ 0.0089 & 0.0252 $\pm$ 0.0045\tabularnewline
                \hline 
                $\| \hat{\Gb} - \Gb^* \|_{F}^{2}/(2r)^{2}$  & 0.0152 $\pm$ 0.0049 & 0.0150 $\pm$ 0.0032 & 0.0343 $\pm$ 0.0125 & 0.0228 $\pm$ 0.0079\tabularnewline
                \hline 
                TPR - fixed & 0.9992 $\pm$ 0.0057 & 0.9988 $\pm$ 0.0048 & 0.9887 $\pm$ 0.0222 & 0.9968 $\pm$ 0.0110\tabularnewline
                \hline 
                FPR - fixed & 0.0154 $\pm$ 0.0141 & 0.0159 $\pm$ 0.0105 & 0.0317 $\pm$ 0.0187 & 0.0231 $\pm$ 0.0178\tabularnewline
                \hline 
                TPR - random & 0.9944 $\pm$ 0.0158 & 0.9946 $\pm$ 0.0126 & 0.9551 $\pm$ 0.0492 & 0.9722 $\pm$ 0.0399\tabularnewline
                \hline 
                FPR - random & 0.0221 $\pm$ 0.0209 & 0.0182 $\pm$ 0.0154 & 0.0465 $\pm$ 0.1005 & 0.0490 $\pm$ 0.0307\tabularnewline
                \hline 
                \% of $\hat{K}=K^*$ & 100\% & 99\% & 94\% & 99\%\tabularnewline
                \hline 
                 & \multicolumn{4}{c}{Alternative univariate response GCM method}\tabularnewline
                \hline 
                $\| \hat{\Bb} - \Bb^* \|_{F}^{2}/pr$ & 0.0430 $\pm$ 0.0098 & 0.0456 $\pm$ 0.0117 & 0.0940 $\pm$ 0.0289 & 0.0585 $\pm$ 0.0119\tabularnewline
                \hline 
                $\| \hat{\Gb} - \Gb^* \|_{F}^{2}/(2r)^{2}$ & 0.1182 $\pm$ 0.0117 & 0.1202 $\pm$ 0.0073 & 0.1190 $\pm$ 0.0117 & 0.1187 $\pm$ 0.0117\tabularnewline
                \bottomrule 
\end{tabular}
}
\end{table}

Table \ref{table-simu-res} reports the results averaged across 100 replications. In terms of the estimation accuracy, it is seen from the table that our proposed method achieves much smaller estimation error for both the fixed effects matrix $\Bb^*$ and the covariance matrix $\Gb^*$ of the random effects compared to the benchmark solution. In addition, the estimation error increases when the sample size $n$ decreases, or when the noise level increases. The estimation error does not change much when the number of outcomes $r$ increases due to the low-rank factorization structure. In terms of the selection accuracy, it is also seen from the table that the proposed method manages to select the true fixed and random slopes, as well as the true rank,  successfully. Again, the selection accuracy decreases as the sample size or the noise level increases, and does not change much as $r$ increases. These observations agree with our expectations. 

\begin{figure}[t!]
\centering
\includegraphics[width=0.49\textwidth]{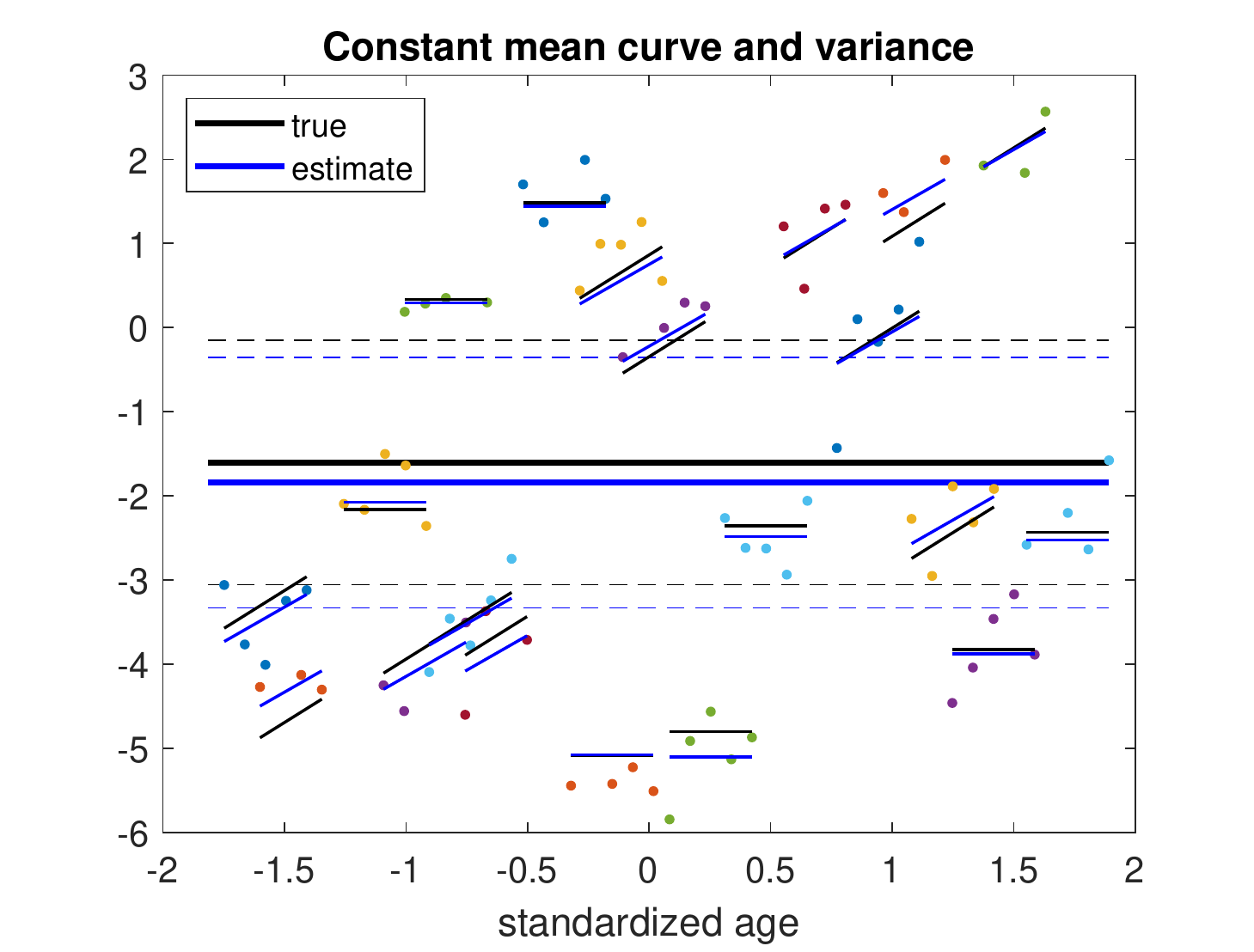}
\includegraphics[width=0.49\textwidth]{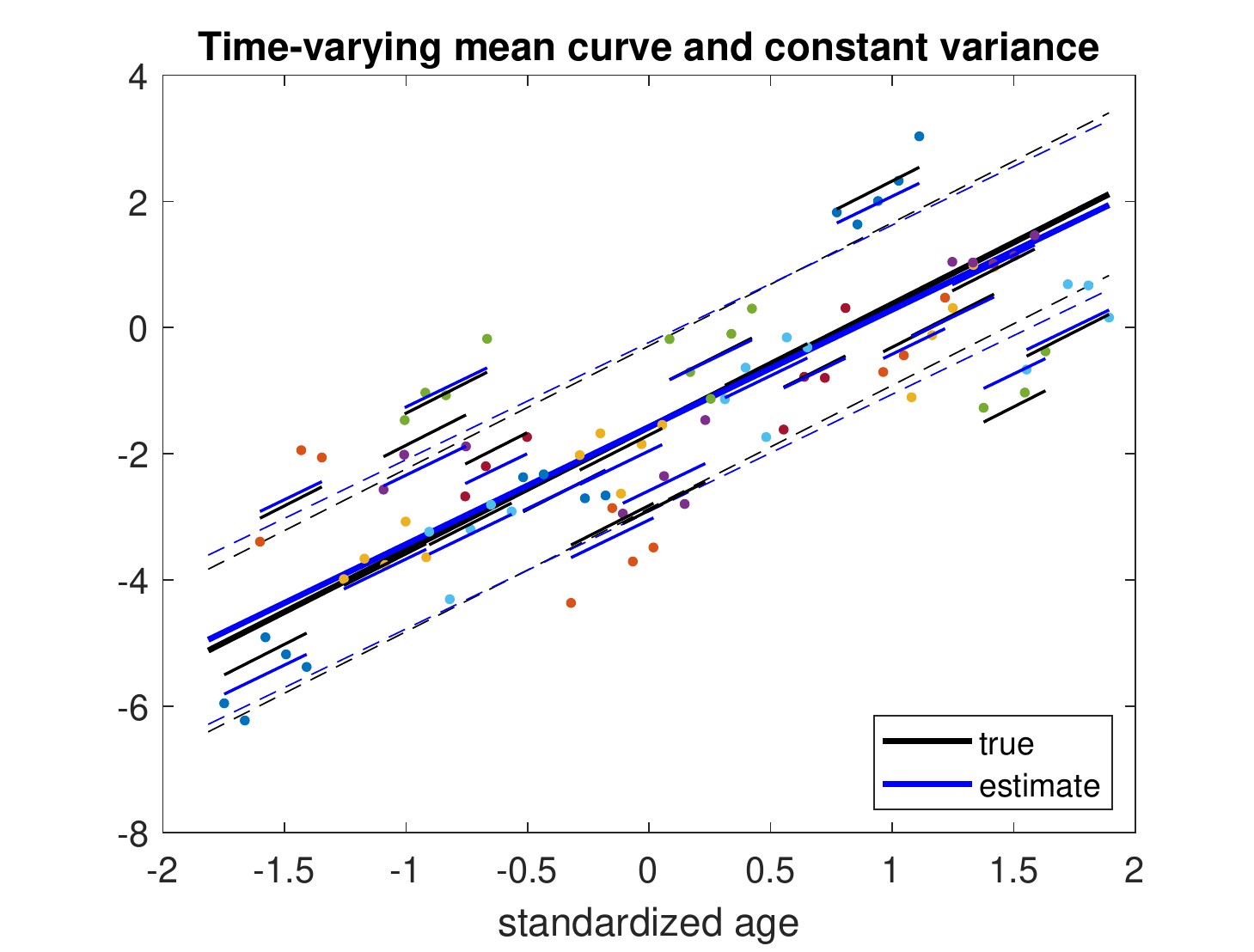} \\
\includegraphics[width=0.49\textwidth]{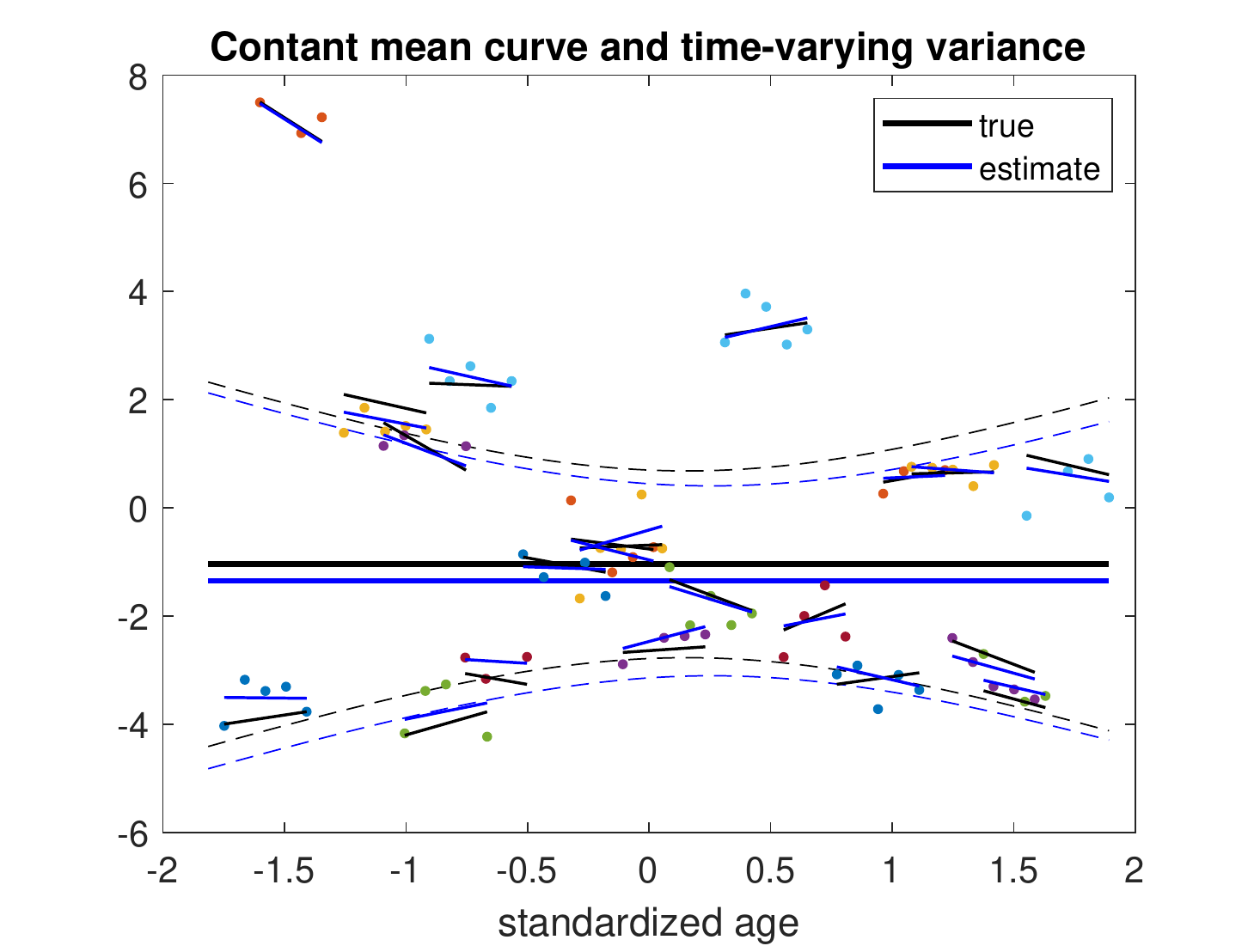}
\includegraphics[width=0.49\textwidth]{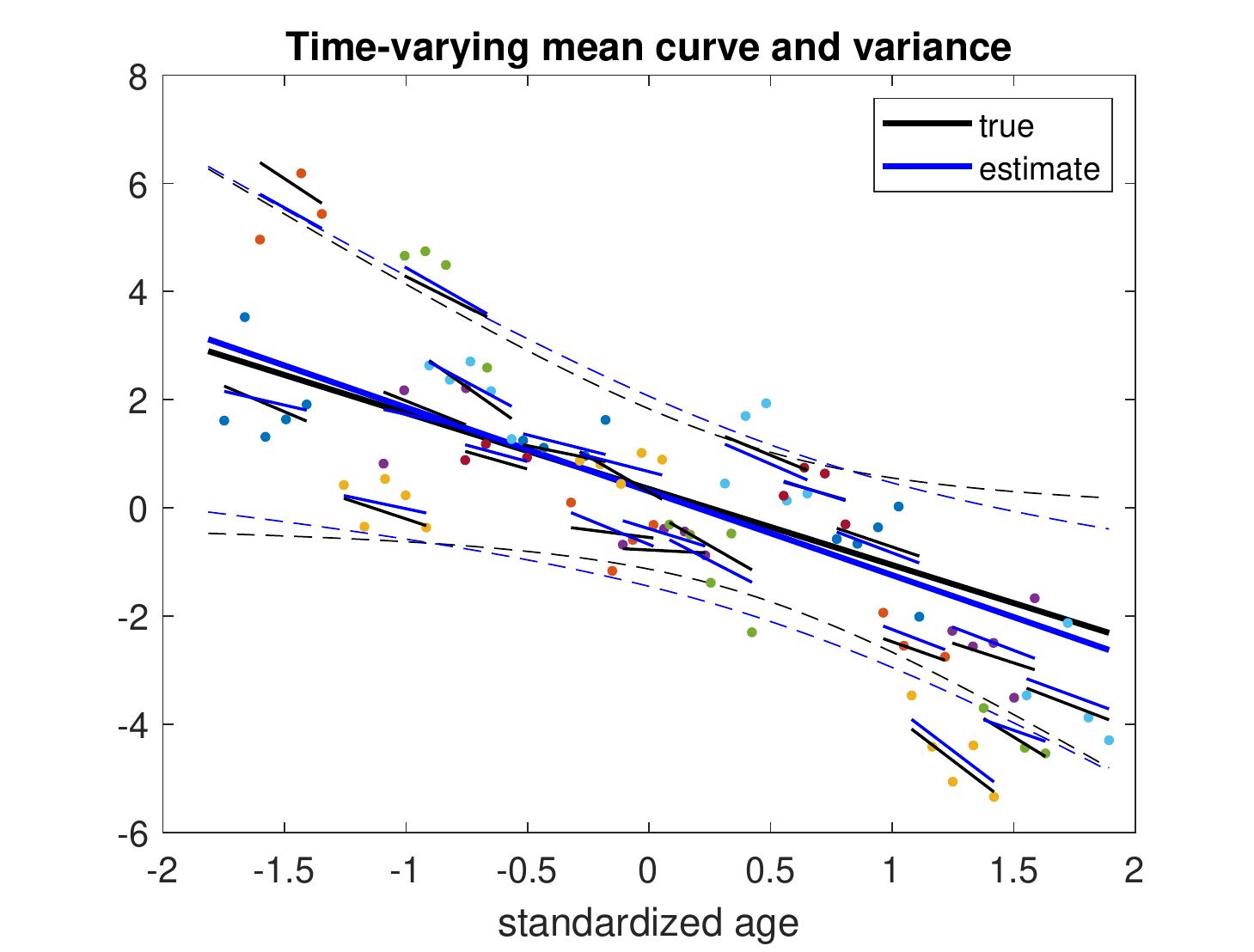}
\caption{Growth curves with respect to age from a single data replication, with $r=100$, $n=100$ and 20\% noise. Shown are the estimated mean growth curves for randomly selected four types of outcomes, along with the individual growth curves of 20 randomly selected subjects. Black lines - the truth; blue lines - the estimated growth curves, bold solid lines - the mean growth curves $(\mu_{0j}+\mu_{1j}g)$, short solid lines - the individual growth curves $(\beta_{0ij}+\beta_{1ij}g_{it})$, dashed lines - the mean growth curves $\pm$ one standard deviation, and dots the observed outcomes for different subjects with different colors.}
\label{fig-4types}
\end{figure}

Figure \ref{fig-4types} reports the results for four types of outcomes from a single data replication. It is seen from the plot that the proposed method correctly recovers the trends of different types of outcomes, and the estimated individual growth curves generally match the truth.

Finally, we report the computational time of our method. All numerical experiments have been conducted on a personal computer with six Intel Core i7 3.2 GHz processors and 64 GB RAM. For the simulation setting with $r=200$ and $n=100$, the proposed method took on average less than 2 minutes for one data replication.

\section{Application}
\label{sec:application}

We illustrate our proposed method using the longitudinal study of brain structural connectivity in association with human immunodeficiency virus (HIV) \citep{tivarusME2021}. The data contains  diffusion tensor imaging (DTI) of 32 HIV infected patients and 60 age-matched healthy controls over a two-year span. The HIV patients were scanned before starting a treatment called combination antiretroviral therapy (cART), then were scanned after 12 weeks, one year, and two years of the cART treatment, respectively. Meanwhile, the healthy subjects were scanned at the beginning of the study, then annually afterwards. Correspondingly, there are $n=92$ subjects, and each subject in total received DTI scans at $T_i = 3$ to $4$ time points. For each DTI scan, the brain structural connectivity is characterized by the connection strength of fiber tracts between pairs of brain regions. We employed the Desikan atlas \citep{Desikan2006968} for brain region parcellation, which includes 68 regions of interest, with 34 in each hemisphere. The connection strength between each region pair is measured by the  fractional anisotropy (FA), a commonly used DTI metric, averaged along all fiber tracts between two brain regions \citep{zhang2018mapping}. After removing the pairs with constant zero connection strength, there are $r = 2006$ connections remained. The chronological age $g_{it}$ ranges between 20 and 71 years old, the time-invariant predictor $\ub_i$ is the binary HIV infection status, with $1$ indicating the infection and $0$ otherwise, and there is no time-varying predictor $\wb_{it}$. The goal is to investigate the dynamic change of brain structural connectivity, and identify brain regions where the connections have different growth behaviors between the HIV patients and healthy controls. We apply the proposed method to the data, and select $\hat{K}=2$ using BIC. 

\begin{figure}[t!]
\centering
\begin{tabular}{cc}
\vspace{-0.2in}
\includegraphics[trim={1.5cm 0 1.5cm 0}, clip, width=0.4\textwidth]{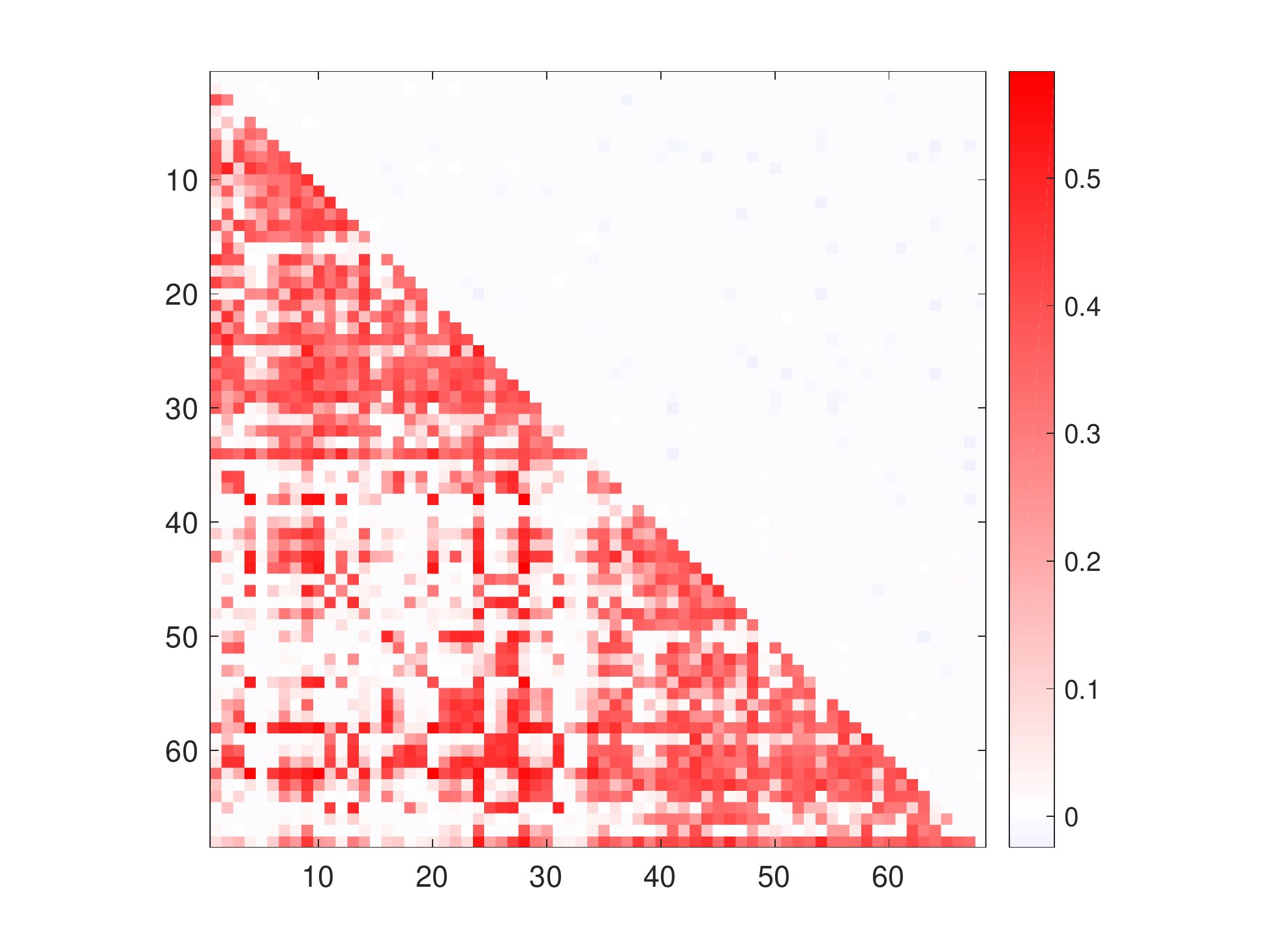} & 
\includegraphics[trim={1.5cm 0 1.5cm 0}, clip, width=0.4\textwidth]{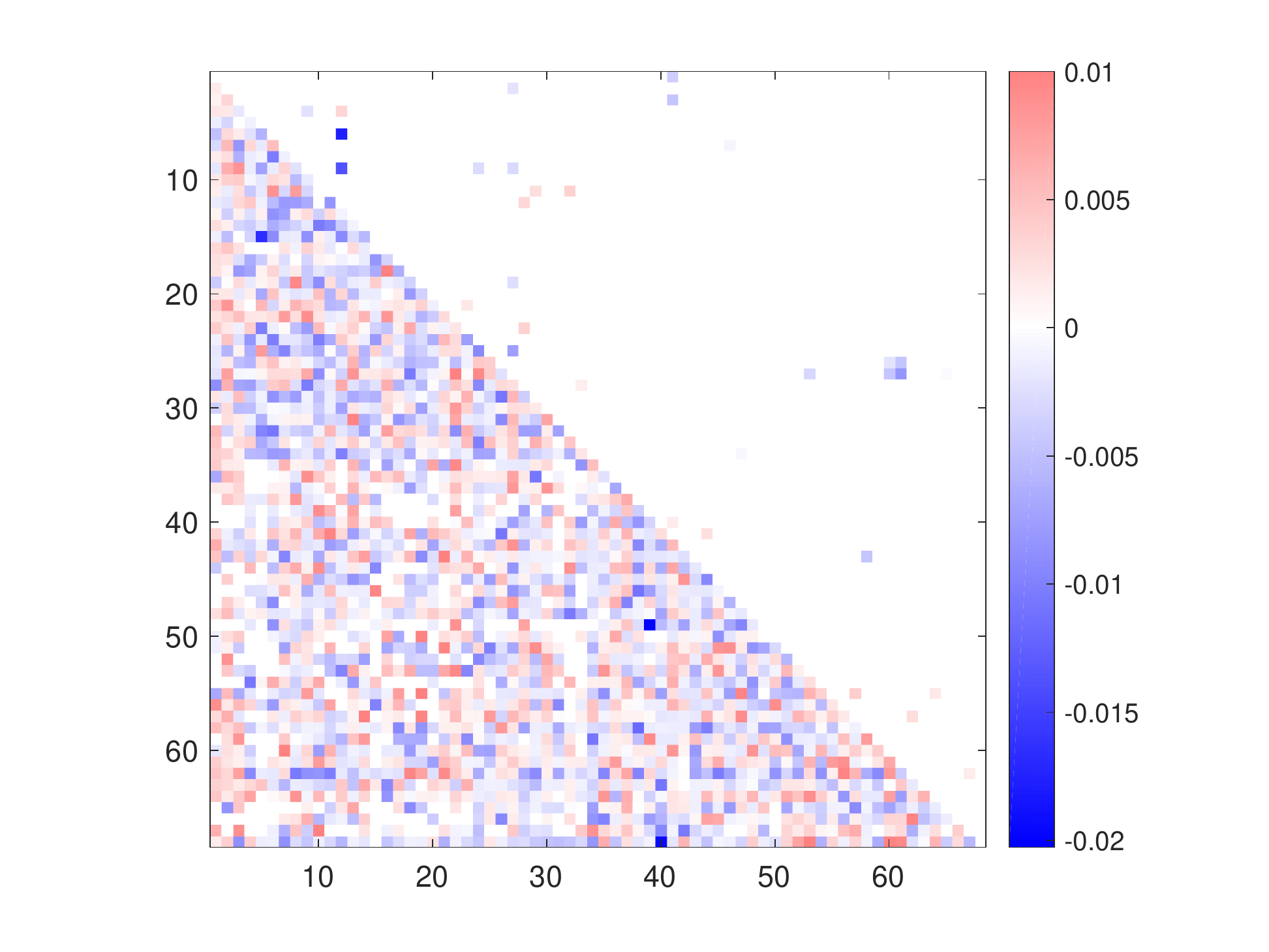} \\
(a) & (b) \\
\vspace{-0.2in}
\includegraphics[trim={1.5cm 0 1.5cm 0}, clip, width=0.4\textwidth]{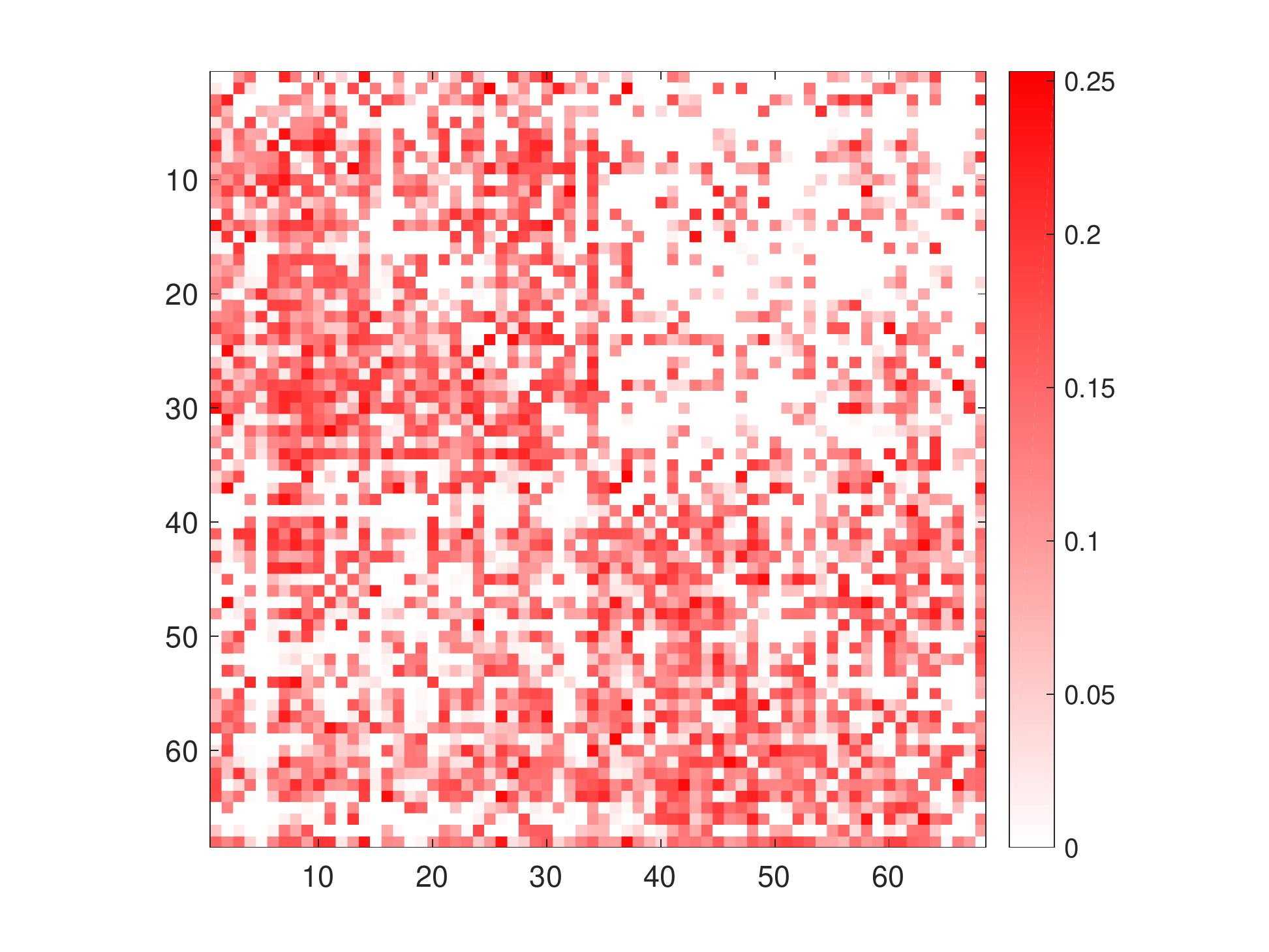} & 
\includegraphics[width=0.475\textwidth]{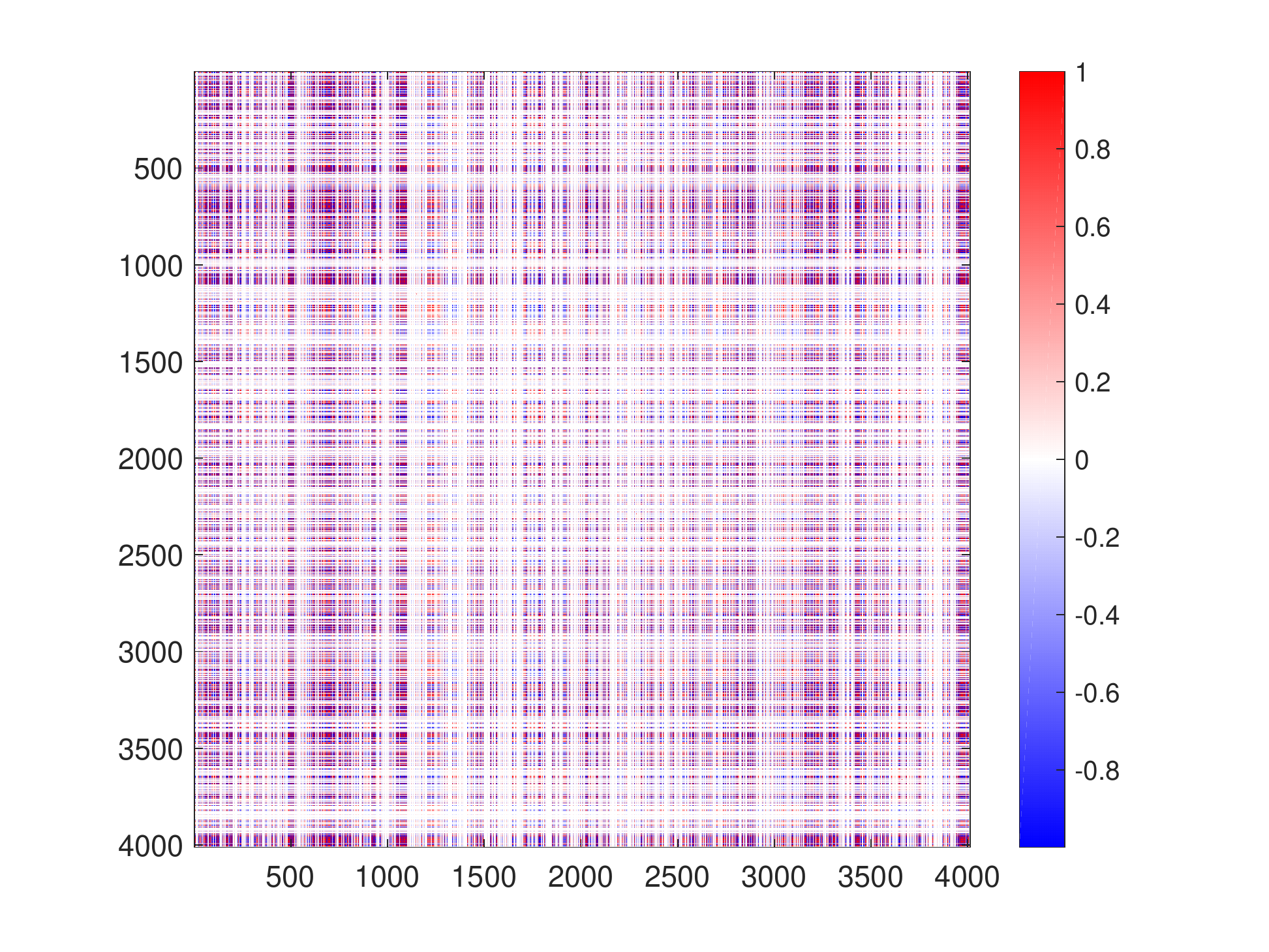} \\
(c) & (d) 
\end{tabular}
\caption{Estimated coefficients for the HIV study of brain structural connectivity. The four panels correspond to (a) $\hat{\mu}_{0j}$ (lower-triangular) and $\hat{\mu}_{1j}$ (upper-triangular); (b) $\hat{\alpha}_{0j}$ (lower-triangular) and $\hat{\alpha}_{1j}$ (upper-triangular); (c) $\hat{d}_{2j-1}$ (lower-triangular) and $\hat{d}_{2j}$ (upper-triangular), $j \in [r]$; and (d) the correlation matrix $\hat{\Rb}$ of all the random effects.}
\label{fig-app-coefficients}
\end{figure}

Figure \ref{fig-app-coefficients} reports the estimated coefficients for all $j \in [r]$. In particular, Figure \ref{fig-app-coefficients}(a) shows the estimated $\hat{\mu}_{0j}$ (lower triangular) and $\hat{\mu}_{1j}$ (upper triangular) that are the intercept and slope of the mean growth curve of healthy subjects. Figure \ref{fig-app-coefficients}(b) shows the estimated $\hat{\alpha}_{0j}$ (lower triangular) and $\hat{\alpha}_{1j}$ (upper triangular) that are the difference in the intercept and in the slope of the mean growth curves between the healthy subjects and HIV patients.  Figure \ref{fig-app-coefficients}(c) shows the estimated $\hat{d}_{2j-1}$ (lower triangular) and $\hat{d}_{2j}$ (upper triangular) that are the standard deviation of the random intercepts $\zeta_{0ij}$ and of the random slopes $\zeta_{1ij}$, $i \in [n]$. Figure \ref{fig-app-coefficients}(d) shows the estimated $2r \times 2r$ correlation matrix $\hat{\Rb}$ of the random effects.

\begin{figure}[t!]
\centering
\begin{tabular}{cc}
\vspace{-0.2in}
\includegraphics[width=0.45\textwidth]{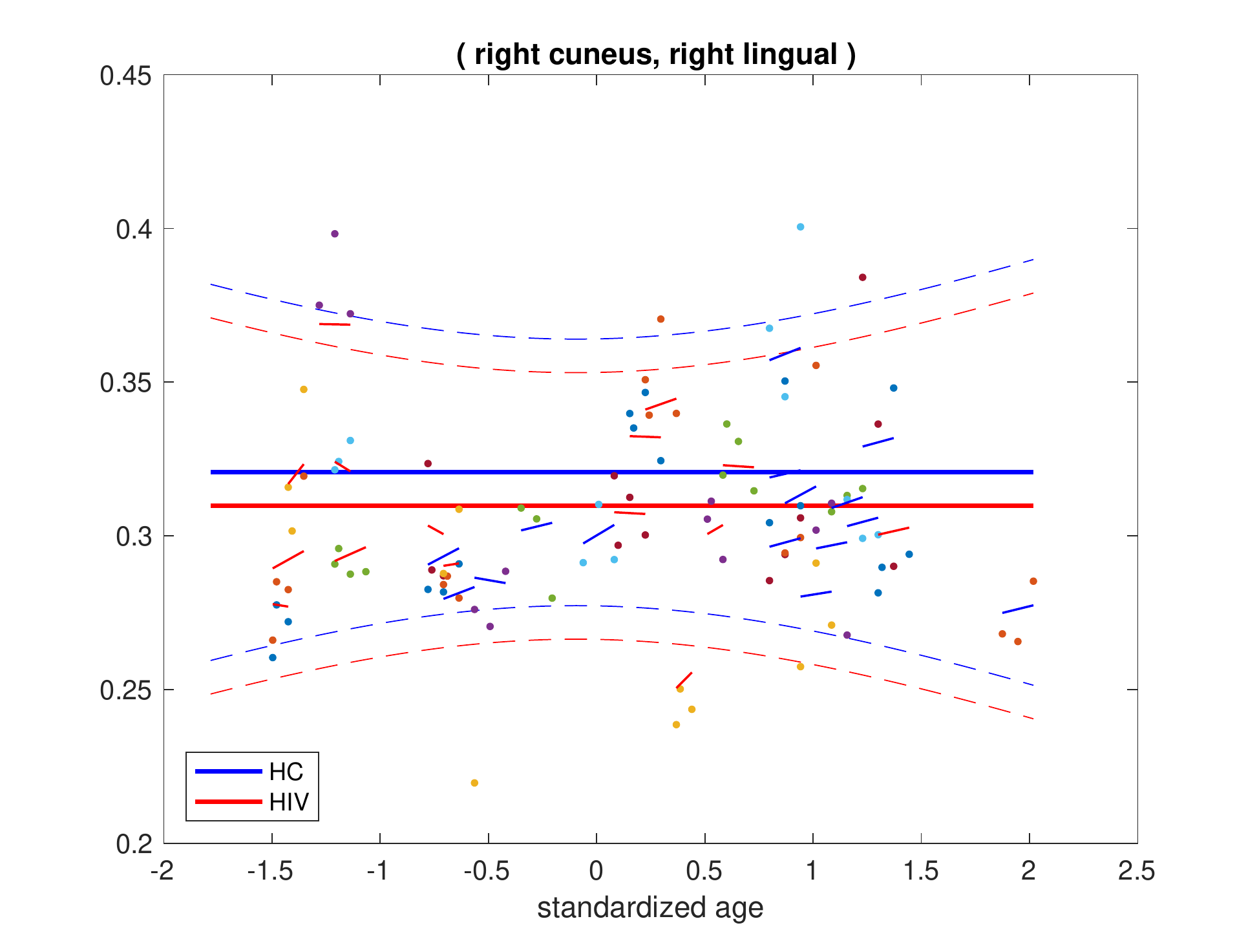} & 
\includegraphics[width=0.45\textwidth]{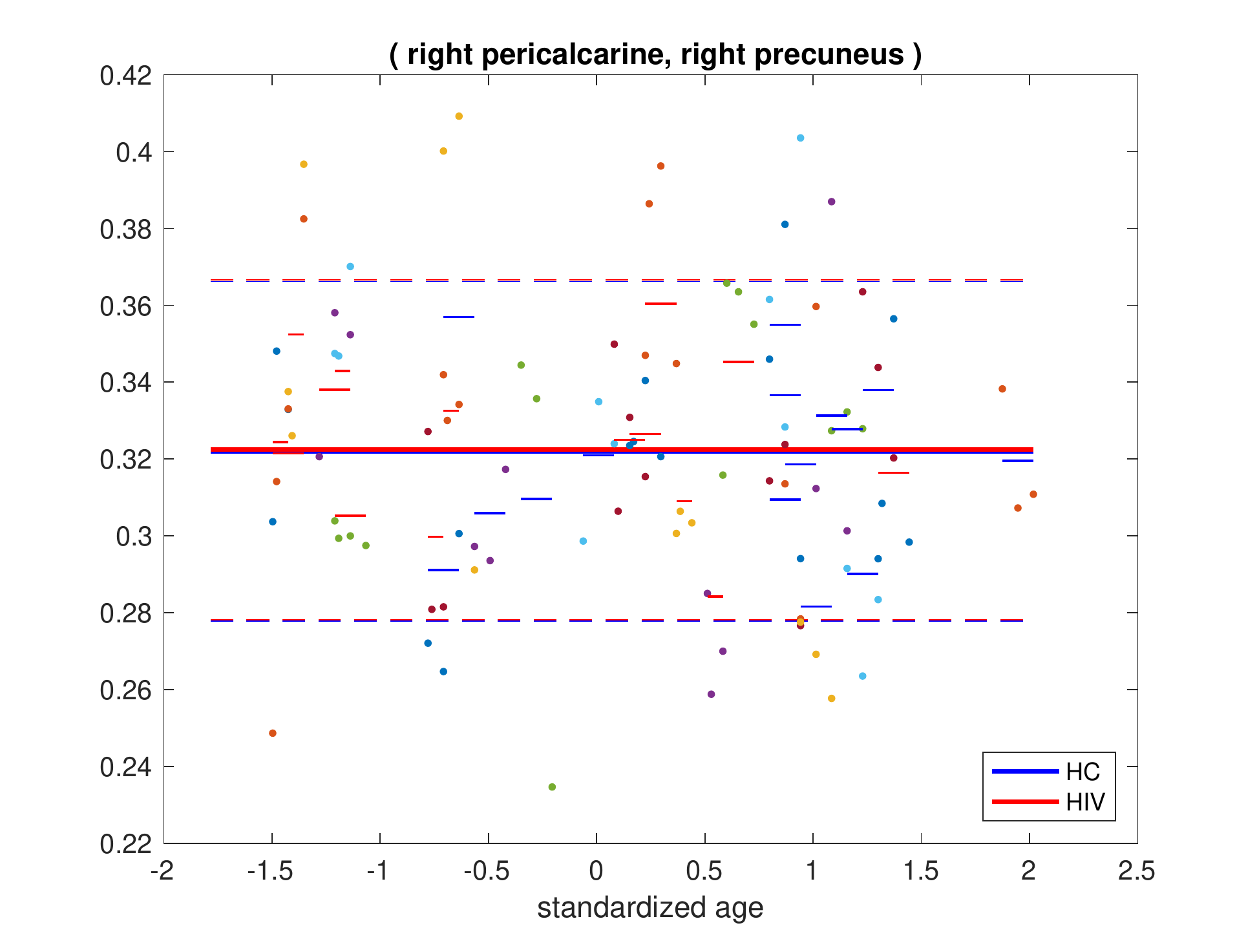}	 \\	
(a) & (b) \\
\vspace{-0.2in}
\includegraphics[width=0.45\textwidth]{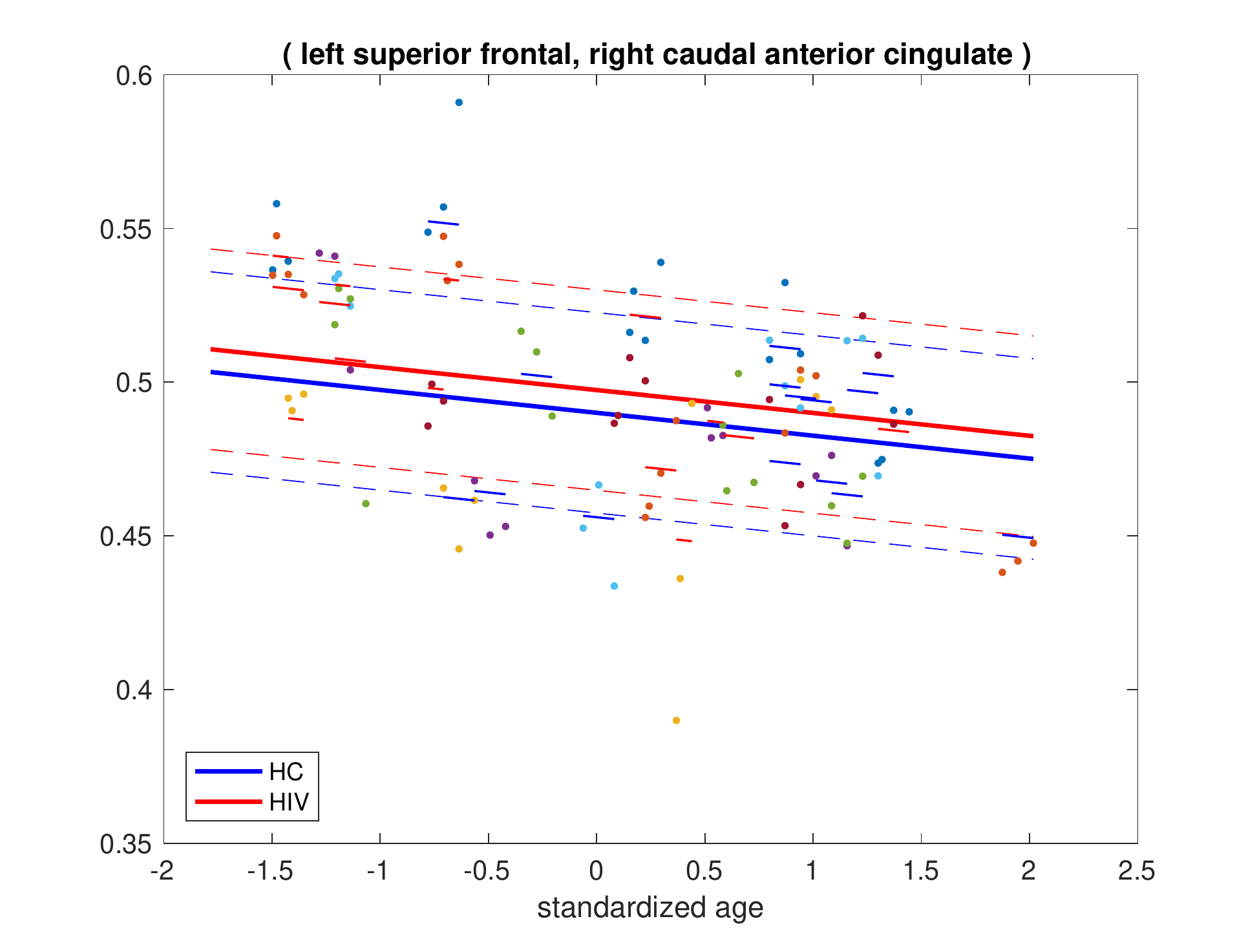} & 
\includegraphics[width=0.45\textwidth]{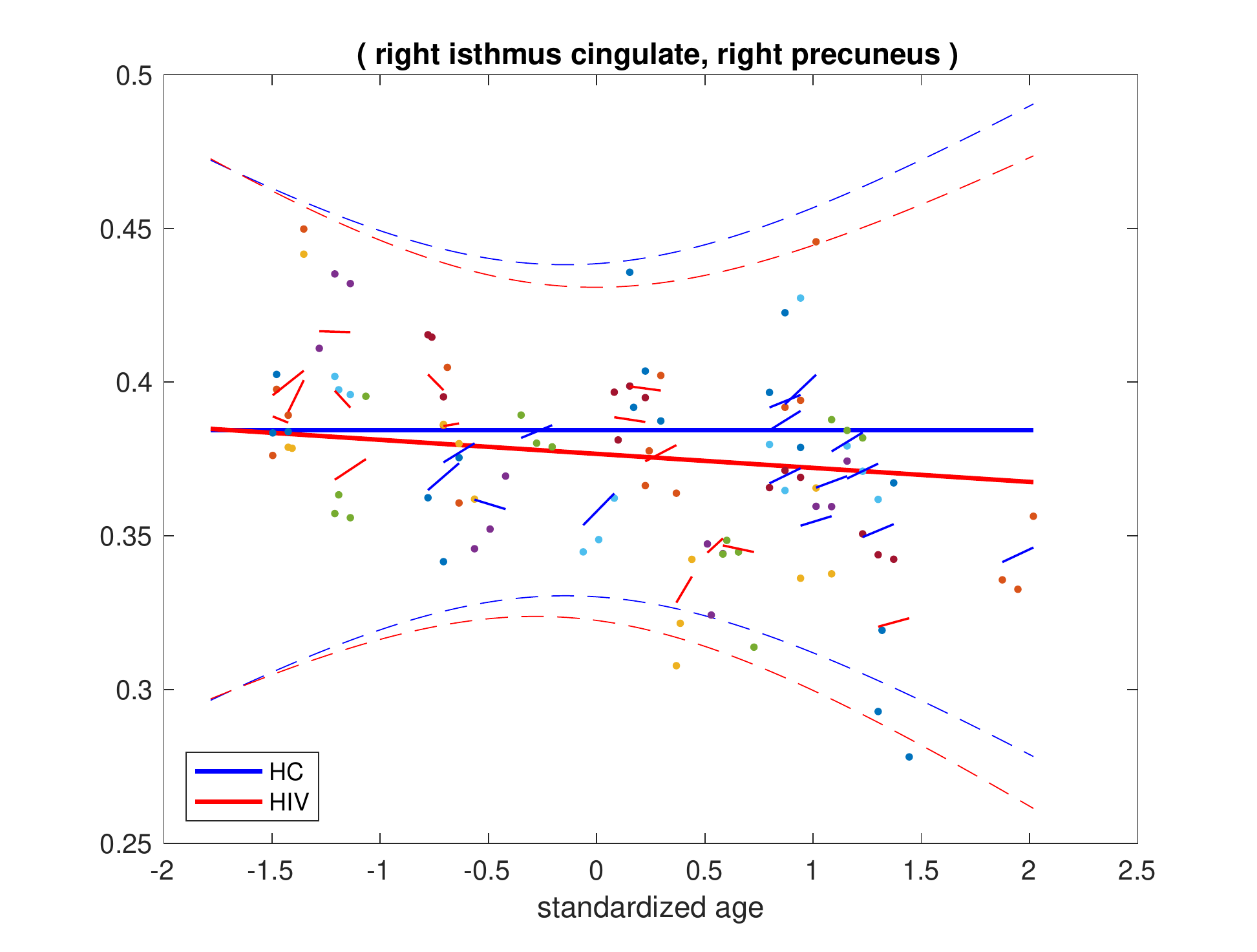} \\
(c) & (d) 
\end{tabular}
\caption{Estimated mean growth patterns for the HIV study of brain structural connectivity. The four panels correspond to four categories of mean growth curves over age for HIV patients (red) and healthy controls (blue), along with individual growth curves of 15 randomly selected HIV patients (red) and 15 controls (blue). The dots denote the observed FA values, and the dashed lines denote the mean growth curves $\pm$ one standard deviation.}
\label{fig-app-4cons}
\end{figure}

Figure \ref{fig-app-4cons} reports the estimated mean growth patterns. Nearly all structural connections can be categorized into four growth patterns, characterized by the combinations of the estimated $\hat{\mu}_{1j}$, $\hat{\alpha}_{1j}$ and $\hat{d}_{2j}$. In particular, Figure \ref{fig-app-4cons}(a) shows a category where $\hat{\mu}_{1j}=0, \hat{\alpha}_{1j}=0$, and $\hat{d}_{2j} \neq 0$. In this case, the mean growth curves of both the HIV patients and controls are constant over age, but the marginal variance of the connection strength is time-varying, and the individual growth curves change at different rates. The plot shows one such connection in this category, where the bold blue line shows the estimated mean growth curve $(\hat{\mu}_{0j}+\hat{\mu}_{1j}g)$ of the healthy subjects, the bold red line shows the estimated mean growth curve $\{ (\hat{\mu}_{0j}+\hat{\alpha}_{0j})+(\hat{\mu}_{1j}+\hat{\alpha}_{1j})g \}$ of the HIV patients, while the short solid lines show the estimated individual growth curves $\ (\hat{\beta}_{0ij}+\hat{\beta}_{1ij}g_{it})$. Figure \ref{fig-app-4cons}(b) shows a category where $\hat{\mu}_{1j}=0, \hat{\alpha}_{1j}=0$, and $\hat{d}_{2j} = 0$. In this case, the mean growth curves of both the HIV patients and controls stay constant over age, the marginal variance of the structural connection is constant, and all individual growth curves are constant over age. Figure \ref{fig-app-4cons}(c) shows a category where $\hat{\mu}_{1j} \neq 0, \hat{\alpha}_{1j}=0$, and $\hat{d}_{2j} = 0$. In this case, the mean growth curves change at the same rate for the HIV patients and controls, whereas the marginal variance of the connection remains constant, and all individual growth curves change at the same rate. Figure \ref{fig-app-4cons}(d) shows a category where $\hat{\mu}_{1j} = 0, \hat{\alpha}_{1j} \neq 0$ and $\hat{d}_{2j} \neq 0$. In this case, the HIV patients have a time-varying mean growth pattern, while the healthy controls have a constant mean growth pattern. In addition, the connection strength has time-varying marginal variance and heterogeneous growth rates across individuals. 

Table \ref{tab-HIV-ROIs} reports the pairs of brain regions with the largest five absolute values of $\hat{\alpha}_{0j}$ and $\hat{\alpha}_{1j}$, since these coefficients indicate large differences in the growth patterns of structural connections between the HIV patients and healthy controls, and are of scientific interest. We note that, for the connections with top absolute values of $\hat{\alpha}_{0j}$, the mean growth curves of the HIV patients and controls are all constant over age. So the negative values of $\hat{\alpha}_{0j}$ imply that the HIV patients have lower connection strengths on average compared to the controls. For the connections with top absolute values of $\hat{\alpha}_{1j}$, the corresponding $\hat{\mu}_{1j} = 0$. So the negative values of $\hat{\alpha}_{1j}$ imply that the HIV patients have decreasing mean growth curves, while the healthy controls have constant mean growth curves. We also remark that, the brain regions identified by our method in Table \ref{tab-HIV-ROIs} are consistent with the current scientific literature. In particular, the lingual gyrus, precuneus, fusiform, and isthmus of the cingulate gyrus have been found to exhibit significantly different microscale brain properties between the HIV-infected subjects and healthy controls \citep{zhuang2021whole}. The insula and parahippocampal gyrus have been shown to have reduced regional volume and functional connectivity in the HIV-infected population \citep{samboju_structural_2018}. Severe depressive symptoms in HIV patients have been associated with loss of hippocampal volume, especially in the entorhinal cortex \citep{bronshteyn2021depression, weber2022}.
The rostral anterior cingulate cortex has shown significant changes in terms of cortical thickness in pediatric HIV patients \citep{yadav2017altered}. The cuneus and superior frontal gyrus have shown significantly lower gyrification index in HIV-infected individuals compared to healthy controls \citep{joy2023alterations}.

\begin{table}[t!]
\centering
\caption{Brain structural connections with 5 largest absolute values of $\hat{\alpha}_{0j}$ and $\hat{\alpha}_{1j}$.}
\label{tab-HIV-ROIs}
\footnotesize{
\begin{tabular}{lc|lc}
\toprule 
Connection & $\hat{\alpha}_{0j}$ & Connection & $\hat{\alpha}_{1j}$\tabularnewline
\midrule
(R.entorhinal, R.parahippocampal) & -0.0203 & (L.fusiform, L.lingual) & -0.0178\tabularnewline
(R.fusiform, R.insula) & -0.0194 & (L.isthmuscingulate, L.lingual) & -0.0136\tabularnewline
(L.entorhinal, L.parahippocampal) & -0.0166 & (L.superiorfrontal, R.superiorfrontal) & -0.0083\tabularnewline
(R.cuneus, R.precuneus) & -0.0118 & (L.rostralanteriorcingulate, L.superiorfrontal) & -0.0076\tabularnewline
(R.cuneus, R.lingual) & -0.0109 & (R.isthmuscingulate, R.precuneus) & -0.0046\tabularnewline
\bottomrule 
\end{tabular}
}
\end{table}

\section{Discussion}
\label{sec:discussion}

In this article, we have mainly considered linear type growth curve models. This is because the observations were collected at merely 3 to 4 time points for each subject, which is common in longitudinal neuroimaging applications. When the data are collected at more time points for each subject, we may consider more flexible growth curve models, for instance, a polynomial model or splines, and we can extend the method developed here in a relatively straightforward fashion. 

In this article, we have focused on the methodological aspect, while the theoretical analysis of our proposed method is very challenging. We note that, to establish the estimation consistency, the current EM theoretical framework relies on two key components, the geometry of the $Q$-function around the true parameters so to quantify the computational error, and the convergence rate of the M-step optimization so to quantify the statistical error \citep{balakrishnan2017statistical}.  In our model, the high-dimensional response and the flexible correlation structures make the likelihood rather complex, which is actually much more complicated than the model studied in the high-dimensional EM regime \citep{cai2019chime}. In addition, our model also involves some special sparsity structure, which further complicates the asymptotic analysis. There is a lack of theoretical tools for our setting, and we leave the theoretical investigation as future research.

\bibliographystyle{Chicago}
\bibliography{ref-longit}

\end{document}